\documentclass[12pt,a4]{article}

\usepackage{amsmath}
\usepackage{amssymb}
\usepackage{epsfig}
\usepackage{graphicx}
\usepackage{cite}
\usepackage{mathrsfs}    
\usepackage{bbm} 

\makeatletter
\renewcommand{\fnum@table}{\textbf{\tablename~\thetable}}
\renewcommand{\fnum@figure}{\textbf{\figurename~\thefigure}}
\makeatother

\newcounter{myenumi}

\renewcommand{\themyenumi}{\roman{myenumi}}
{\end{list}}

\newlength{\myem}
\newcounter{mysubequation}[equation]

\makeatletter
\renewcommand{\section}{\@startsection{section}{1}{0em}{-\baselineskip}%
{\baselineskip}{\normalfont\large\bfseries}}
\renewcommand{\subsection}%
{\@startsection{subsection}{2}{0em}{-0.7\baselineskip}%
{0.7\baselineskip}{\normalfont\bfseries}}
\makeatother

\newcommand{\SM}{\mathrm{SM}}
\newcommand{\bulk}{\mathrm{bulk}}

\newcommand{\hc}{\mathrm{h.c.}}
\newcommand{\MPl}{M_{\textrm{Pl}}}

\newcommand{\diag}{\mathrm{diag}}

\newcommand{\eff}{_\textrm{eff}}
\newcommand{\weak}{_\textrm{eff,w}}
\newcommand{\strong}{_\textrm{eff,s}}
\newcommand{\branebulk}{\mathrm{brane-bulk}}

\newcommand{\I}{\mathrm{i}}

\newcommand{\Gr}{\mathrm{Gr}}
\newcommand{\eq}{^\textrm{eq}}

\begin{document}

\begin{titlepage}

\renewcommand{\thefootnote}{\alph{footnote}}

\vspace*{-3.cm}
\begin{flushright}
TUM-HEP-669/07
\end{flushright}

\vspace*{0.5cm}

\renewcommand{\thefootnote}{\fnsymbol{footnote}}
\setcounter{footnote}{-1}

{\begin{center}
{\Large\bf Leptogenesis With Many Neutrinos}
\end{center}}

\vspace*{.8cm}
{\begin{center} {\large{\sc
                Marc-Thomas Eisele\footnote[1]{\makebox[1.cm]{Email:}
                eisele@ph.tum.de}}}
\end{center}}
\vspace*{0cm}
{\it
\begin{center}

\vspace*{1mm}

       Physik--Department, Technische Universit\"at M\"unchen, \\
       James--Franck--Strasse, 85748 Garching, Germany
\vspace*{1mm}

\today
\end{center}}

\vspace*{0.3cm}

\begin{abstract}
We consider leptogenesis in scenarios with many neutrino singlets. We find that the lower bound for the reheating temperature can be significantly relaxed with respect to the hierarchical three neutrino case. We further argue that the upper bound for the neutrino mass scale from leptogenesis gets significantly lifted in these scenarios.
As a specific realization, we then discuss an extra-dimensional model, where the large number of neutrinos is provided by Kaluza-Klein excitations.
\end{abstract}

\vspace*{.5cm}

\end{titlepage}

\newpage

\renewcommand{\thefootnote}{\arabic{footnote}}
\setcounter{footnote}{0}


\section{Introduction}

Among the many open issues within astro-particle physics and cosmology, the origin of the observed baryon asymmetry of the universe (BAU) \cite{Spergel:2003cb}
\begin{equation} \label{eq measured asym}
	\eta _B \equiv \frac {n_B}{n_\gamma} 
	= 6.1^{+0.3}_{-0.2} \times 10^{-10} \, 
	\end{equation}
is one of the most frequently discussed, as it cannot be explained within the standard model of particle physics (SM). Here, $n_B\equiv  n_b- n_{\bar{b}}$ is the difference of today's baryon and anti-baryon densities and $n_\gamma$ is the corresponding value for the photons.

Especially, in the context of inflation, which can solve many cosmological problems, such as the flatness and the horizon problem, one typically starts with an empty universe at the end of this epoch, which implies that somehow a baryon-asymmetry needs to be produced before big bang nucleosynthesis (BBN) sets in.

One of the most popular models that can produce this asymmetry is leptogenesis \cite{Fukugita:1986hr,Luty:1992un,Plumacher:1996kc,Plumacher:1997ru,Barbieri:1999ma,Buchmuller:2002rq,Giudice:2003jh,Buchmuller:2004nz}. The popularity of this model can be ascribed to the fact that it does not need any additions to the SM other than right-handed neutrinos, which are also needed to make the SM neutrinos massive, as indicated by their observed oscillations. If these right-handed neutrino states are very heavy, they can nicely explain the smallness of the observed neutrino masses via the see-saw mechanism. In leptogenesis, these state are indeed assumed to be heavy. After their excitations in the early universe they decay in a CP violating manner, leaving a lepton asymmetry, which is then partially transformed to a baryon asymmetry by so called sphaleron processes \cite{Klinkhamer:1984di,Kuzmin:1985mm}. Lately, much attention has also been given to possible flavor effects in this scenario \cite{Barbieri:1999ma,Pilaftsis:2003gt,Abada:2006fw,Nardi:2006fx,Abada:2006ea}.

However, the typical lower bound for the reheating temperature $T_r \gtrsim 10^9$GeV \cite{Davidson:2002qv,Giudice:2003jh,Buchmuller:2004nz}, which is needed to produce a large enough baryon asymmetry via leptogenesis is only marginally compatible with the upper bound on the same value in supergravity theories $T_r<10^9-10^{10}$GeV \cite{Khlopov:1984pf,Ellis:1984eq}, where an exceeding of this temperature would alter the predictions of BBN due to an overproduction of gravitinos. Bounds for more constrained models can be found in Refs. \cite{Moroi:1993mb,Khlopov:1993ye,Asaka:2000zh,Roszkowski:2004jd,Cerdeno:2005eu,Steffen:2006hw,Pradler:2006hh}.

Several ways to overcome this discrepancy have been considered, such as resonant leptogenesis \cite{Pilaftsis:2003gt} and soft leptogenesis \cite{Grossman:2003jv,D'Ambrosio:2003wy}. In the present approach, we consider the effect of many heavy right-handed neutrinos (i.e. SM gauge singlets) on standard thermal leptogenesis in the one-flavor approximation. We find that a large number of heavy singlet states in the weak wash-out regime can significantly lower the required reheating temperature without any resonant enhancement of the CP asymmetries. For the strong wash-out regime we find a slight decrease of the final baryon asymmetry. While the strong wash-out scenario is typically considered as more natural in the standard case, we argue that the weak wash-out regime can be more natural in models with many singlets.  We also consider the limits of this scenario, which are reached once the heavy singlets come to dominate the energy density of the universe.

Extensions of the standard model with many singlets can be motivated in several ways. Here, we focus on an extra-dimensional model that includes Kaluza-Klein excitations of additional singlet states. These excitations can be considered as right-handed neutrinos. The model
is found to naturally bring all the ingredients that are needed to lower the required reheating temperature within the present approach. Other motivations for setups with many singlet neutrinos come from string models \cite{Buchmuller:2006ik,Buchmuller:2007zd}.

The structure of the paper is as follows. After a review of standard thermal leptogenesis in section 2, where we also derive some rule-of-thumb formulae for the final asymmetries and present parameter bounds for the scenario, we consider possible effects of many heavy singlet states on leptogenesis in section 3. Here, the simplified considerations from the previous section will help us to understand the effects of the new particles, while we also present numerical results that support our considerations. In section 4, we present the already mentioned extra-dimensional scenario that contains all the required ingredients to yield the observed BAU at lower reheating temperatures. We also consider bounds for this scenario from production and decay of extra-dimensional graviton modes. In section 5, we then conclude with a brief summary of our results and a short outlook on possible continuations of the present work. In Appendix A, we present the reaction rates used for the numerical calculations presented throughout the paper, whereas we provide the reader with technical background on the Davidson-Ibarra bound in the presence of many neutrinos in Appendix B.


\section{Standard Thermal Leptogenesis}

In this section we review the standard thermal leptogenesis case \cite{Fukugita:1986hr,Luty:1992un,Plumacher:1996kc,Plumacher:1997ru,Barbieri:1999ma,Buchmuller:2002rq,Giudice:2003jh,Buchmuller:2004nz}, with three hierarchical heavy neutrinos in the one flavor approximation.

Therefore, we need to add three gauge singlets to the standard model, which are assumed to have Majorana mass terms as well as Yukawa interactions with the SM lepton doublets
     \begin{equation} \label{eq singlet addition to L} 
                {\cal L}= -\frac 12 M_i \bar N_i N_i
                - g_{ij} \bar N_i \ell _j \phi  + \mbox{h.c.}\, .
                \end{equation} 
Here, we work in a basis in which the Majorana mass matrix is diagonal, while we assume $M_i \gg g_{jk} \langle \phi \rangle$ with $\langle \phi \rangle$ being the vacuum expectation value (VEV) of the SM Higgs.

\begin{figure}[t]
	\begin{center}\includegraphics[%
	  width=0.4\columnwidth,keepaspectratio]
	  {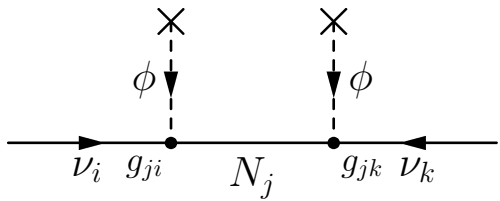}
	  \end{center}
	\caption{\label{fig see-saw}\textit{This Feynman graph illustrates the
	 well-known see-saw mechanism, where the large masses $M_j$ in the
	  propagator of $N_j$ and the VEV of the SM Higgs field lead to an
	  effective Majorana mass matrix for the light neutrinos.}}
	\end{figure}

After electro-weak symmetry breaking, the light (SM) neutrinos also gain an effective Majorana mass matrix via the well-known see-saw mechanism, with the three eigenvalues $m_1$, $m_2$, and $m_3$. To lowest order in $\langle \phi \rangle /M_i$ this can be illustrated by the Feynman graph in fig.\ref{fig see-saw}.

\begin{figure}
	\begin{center}\includegraphics[%
	  width=0.3\columnwidth,keepaspectratio]
	  {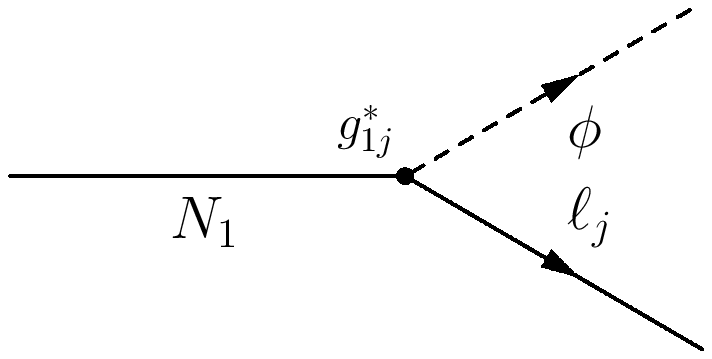}\hspace{1ex}
	  \includegraphics[%
	  width=0.3\columnwidth,keepaspectratio]
	  {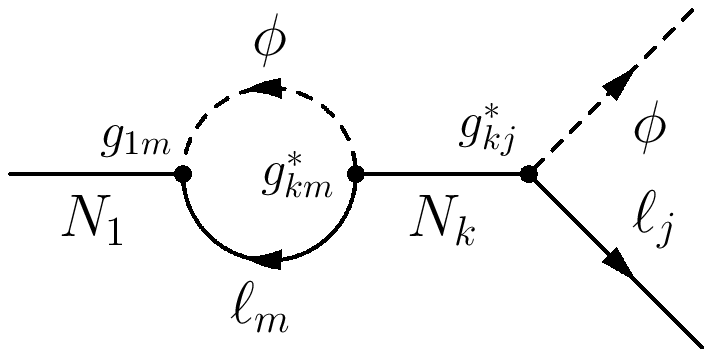}\hspace{1ex}\vspace*{2ex}
	  \includegraphics[%
	  width=0.3\columnwidth,keepaspectratio]
	  {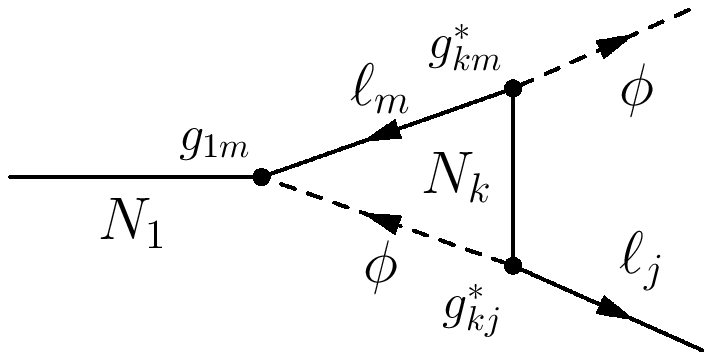}
	  \end{center}
	\caption{\label{fig lept graphs}\textit{The relevant decay-diagrams for standard thermal leptogenesis, where the interference of the first diagram with each of the other two can lead to CP violation and a net lepton number production per decay. Due to the different projection operators, the wave function correction with charge flow in the opposite direction only contributes at higher orders of $M_1/M_k$.}}
	\end{figure}

The addition of the singlets also opens new possibilities for CP-violation, e.g. in the decay of the heavy right-handed neutrinos, where an interference of tree-level and loop diagrams (cf. fig.\ref{fig lept graphs}) can lead to different decay rates for the channels $N \rightarrow \ell \phi ^*$ and $N \rightarrow \bar \ell \phi $. Namely, in the case of hierarchical heavy neutrinos ($M_1 \ll M_2 \ll M_3$) one finds for the decay of the lightest of them \cite{Covi:1996wh}

\begin{equation} \label{eq CP standard}
	\varepsilon _1= \frac {\Gamma(N_1 \rightarrow  \ell_L + \Phi)-
	  	\Gamma(N_1 \rightarrow  \bar \ell_L + \Phi ^*)}
		{\Gamma(N_1 \rightarrow  \ell_L + \Phi)+
	  	\Gamma(N_1 \rightarrow  \bar \ell_L + \Phi ^*)}
		\approx - \frac 3{16}
		\sum _k \frac {\textrm{Im}\left[ (g g^\dagger)^2_{1k} \right]}
		{(g g^\dagger)_{11}} \, \displaystyle \frac {M_1}{M_k} \, .
	\end{equation}
	
One can show that there exists an upper bound for this asymmetry \cite{Hamaguchi:2001gw,Davidson:2002qv}, which is called the Davidson-Ibarra bound and which is given by
\begin{equation}\label{eq davidson ibarra more general}
	|\varepsilon _1| \lesssim 
		\, \frac{3 M_1}
		{16 \pi \langle \phi  \rangle ^2
	 }(m_3-m_1)\,  =  \, 
		\frac{3 M_1 \Delta m^2_\textrm{atm}}
		{16 \pi \langle \phi  \rangle ^2
	 (m_3+m_1)} \, ,
	\end{equation}
that is maximized by $m_1=0$. In this case we have $\Delta m^2_\textrm{atm}=m_3^2$ and
\begin{equation} \label{eq davidson ibarra}
	|\varepsilon _1| \lesssim 
	 \frac 3{16\pi} \frac{M_1 m_3}{\langle \phi \rangle ^2}\approx 
	  10^{-7}\left(\frac {m_3}{0.05 \textrm{eV}} \right) 
	   \left(\frac {M_1}{10^9 \textrm{GeV}} \right) \, 
	   \end{equation}
with almost no freedom left for the size of $m_3$.

As pointed out in ref. \cite{Abada:2006ea} the possible CP-violation within $\Delta L=1$ scatterings can also be relevant for leptogenesis. There, it is also shown that the CP-asymmetry produced in these scatterings exactly equals the asymmetry of the corresponding decay processes. 
	   
At the beginning of the scenario, we assume to be in a homogenous and isotropic universe (possibly after an inflationary period) dominated by the SM particles, which are all in thermal equilibrium with temperature $T_r$. The initial abundance of the right-handed neutrinos is typically assumed to be zero. 

The abundance of right-handed neutrinos at later times is then determined by the differential Boltzmann equation \cite{Kolb:1979qa,Luty:1992un}

\begin{equation} \label{eq boltzmann standard N}
	\frac {dN_{N_1}}{dz} = - \frac {1}{Hz} 
	(\Gamma _D + \Gamma _{\Delta L =1}) \, 
	(N_{N_1}-N^\textrm{eq}_{N_1}) \, ,
	\end{equation}
where $z\equiv M_1/T$ is effectively a reparametrization of time in a background dominated universe. Further, $N_{N_1}$ is the average abundance of the lightest right-handed neutrino in a comoving volume containing one photon at early times, $N^\textrm{eq}_{N_1}$ is the corresponding value for the equilibrium distribution, $\Gamma _D$ is their decay rate, and $\Gamma _{\Delta L =1}$ the rate for the $2\leftrightarrow 2$ scattering processes, which violate lepton number by one (e.g. $N_1 \ell \leftrightarrow t q$). The explicit values for these quantities can be found in appendix \ref{sec app reac rates}.

The second important equation in this context is the one for the $B-L$-asymmetry $N_{B-L}$ \cite{Kolb:1979qa,Luty:1992un,Abada:2006ea} in the same volume
\begin{equation} \label{eq boltzmann standard B-L}
	\frac {dN_{B-L}}{dz} = - \frac {1}{Hz} \left[
	\varepsilon _1 (\Gamma _D + \Gamma _{\Delta L =1})\, 
	(N_{N_1}-N^\textrm{eq}_{N_1})
	+ \Gamma _W N_{B-L} \right] \, ,
	\end{equation}
where $\Gamma _W $ is the reaction rate for wash-out processes consisting of inverse decays and \mbox{$\Delta L =1,2$} scattering. The explicit formulae of the wash-out rates can also be found in appendix \ref{sec app reac rates}.

A fraction of the final $B-L$ asymmetry will be made up of baryons due to sphaleron processes that partially convert a lepton asymmetry into a baryon asymmetry ($N_B=28/79 N_{B-L}$ \cite{Khlebnikov:1988sr,Harvey:1990qw}). To find the final baryon asymmetry per photon $\eta _B$ in a late universe, one also has to consider the late time production of photons, which leads to a dilution of the asymmetry \cite{Buchmuller:2004nz}:
\begin{equation} \label{eq dilution factor}
	\eta _B \approx 10^{-2}N_{B-L}(t\rightarrow \infty) \, .
	\end{equation}
	
Extensive numerical and analytical analysis of equations (\ref{eq boltzmann standard N}) and (\ref{eq boltzmann standard B-L}) have been performed in refs. \cite{Fukugita:1986hr,Luty:1992un,Plumacher:1996kc,Plumacher:1997ru,Barbieri:1999ma,Buchmuller:2002rq,Giudice:2003jh,Buchmuller:2004nz,Abada:2006ea}. In this part of the paper, we aim at understanding the behavior of the solutions of these equations qualitatively to conceive the effect of many decaying neutrinos in the following sections. For reasons of simplicity we will adopt the treatment of ref. \cite{Buchmuller:2004nz} and restrict our considerations to decays, inverse decays and $\Delta L=1$ scatterings involving the top quark. We will comment on $\Delta L=2$ scatterings and scatterings involving gauge bosons later in this section.

When analyzing the Boltzmann equations of leptogenesis, it is useful to define the parameter 
\begin{equation}
	K=\frac {\Gamma _D(z\rightarrow \infty)}{H(z=1)} \, ,
	\end{equation}
which gives a measure for the decoupling of the right-handed neutrinos at the time when their equilibrium abundances start being suppressed. We distinguish the weak wash-out regime ($K\ll 1$) and the strong wash-out regime ($K \gg 1$) and will discuss both of them in the following subsections.

However, before going into detail let us note that the parameter $K$ is often characterized by the relation
\begin{equation} \label{eq K def}
	K\equiv \frac {\tilde m_1}{m_*} \, ,
	\end{equation}
where
\begin{equation} \label{eq eff nu mass 4d}
	\tilde m_1 \equiv \frac {(m_D m_D^\dagger)_{11}}{M_1}
	\end{equation}
is the so-called effective neutrino mass and
\begin{equation}
	m_*\equiv \frac {16 \pi ^{\frac 52}\sqrt{g_*}}{3 \sqrt{5}} 
		\frac {\langle \phi ^2 \rangle}{\MPl}
		\approx 10^{-3} \textrm{eV} 
	\end{equation}
is the called the equilibrium neutrino mass, with the Planck mass $\MPl \equiv 1/G^{1/2}$ and $G$ being Newton's constant.

\subsection{The weak wash-out scenario} \label{sec The weak wash-out scenario} 

In the weak wash-out scenario ($K\ll 1$) the abundance of right-handed neutrinos is never comparable to that of photons and other species due to their weak couplings. For the same reason the wash-out processes, which always involve right-handed neutrinos, never come into thermal equilibrium (i.e. $\Gamma _W \ll H$ at all times). 

However, since decays and inverse decays (as well as the corresponding $\Delta L=1$ scatterings) create $B-L$ asymmetries of opposite sign but equal magnitude, most of the $B-L$ asymmetry which gets produced during the decay of the heavy neutrinos only cancels an asymmetry that was produced during the population of these states. Therefore, even though the wash-out processes are suppressed they are still crucial.

Let us make some rough estimations for the reaction rates based on dimensional analysis.
For $T\gtrsim M_1$ we approximate \cite{Kolb:1979qa}
\begin{eqnarray}
	\Gamma _D &\approx& \Gamma _{ID} \approx	
			\frac 1{8 \pi} (gg^\dagger)_{11}M_1 \frac {M_1}T  \\
	\Gamma _{\Delta L=1} &\approx& 
			\frac{|g_{\textrm{top}}|^2}{\pi ^3}\,
			 (gg^\dagger)_{11}\, T  
			\stackrel {z\sim 1}{\approx }\frac 12 \Gamma _D
			\label{eq approx gamma L=1} 
	\end{eqnarray}
where $g_{\textrm{top}} $ is the Yukawa coupling constant of the top quark. 

Since the right-handed neutrinos will never reach equilibrium abundance until this value becomes strongly suppressed we can put $N^\textrm{eq}_{N_1}-N_{N_1}\approx N^\textrm{eq}_{N_1}$ for $z \lesssim 1$. Using eq.(\ref{eq boltzmann standard N}) we can therefore approximate 
\begin{equation} \label{eq abund N weak}
	N_{N_1}(z\approx 1) 
		\approx
		\frac {(\Gamma _D + \Gamma _{\Delta L =1})N^\textrm{eq}_{N_1}}
		{H(z=1)} \approx  K \,
	\end{equation}	
with $N^\textrm{eq}_{N_1}=3/4$ (at early times).

Without wash-out effects, the $B-L$ asymmetry around this time would be
$N_{B-L}(z=1) \approx - \varepsilon _1 N_{N_1}(z\approx 1)$. However a small fraction of this has been washed-out. To find the size of this fraction, we further approximate
\begin{equation} \label{eq approx Gamma W weak-wo}
	\Gamma _W \approx \frac 12 \Gamma _{ID} + \frac 23 \Gamma _ {\Delta L=1}
	\approx \Gamma _D
	\, , 
	\end{equation}
where the factor $1/2$ is due to fact that half of the inverse decays will finally end up as the original states and the factor $2/3$ is due to the fact that the s-channel process is suppressed more strongly in the weak wash-out before $z\approx 1$ \cite{Luty:1992un}.

Using eq.(\ref{eq boltzmann standard B-L}) this gives as the very rough estimate (cf. \cite{Buchmuller:2004nz})
\begin{eqnarray}
	N_{B-L}(z=1) &\approx & \varepsilon _1 N_{N_1}(z\approx 1)
	\exp \left(- \frac {\Gamma _W}{H(z=1)} \right) \\
	&\approx &  \varepsilon _1   K 
		\exp \left(- K  \right) \\
	&\approx &  \varepsilon _1   K 
		\left( 1-K \right) \, . \label{eq b-l approx z=1}
	\end{eqnarray}

\begin{figure}
	\begin{center}\includegraphics[%
	  width=0.5\columnwidth,keepaspectratio]
	  {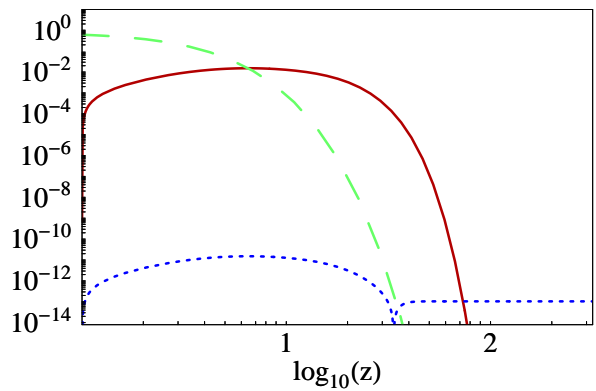}
	  \end{center}
	\caption{\label{fig weak wash-out 1 part}\textit{A typical example for a weak wash-out scenario. The green/dashed line shows the equilibrium density of the right-handed neutrinos $N_1$ per relativistic photon, while the red/solid line shows their actual density as calculated from the Boltzmann equations. The blue/dotted line shows the absolute amount of the $B-L$ density. The graph was plotted with the parameters $K=10^{-2}$, $M_1=10^7$ GeV, $T_r=M_1$, $\varepsilon _1= 10^{-9}$, and a Higgs boson mass of $10^{-1} M_1$. One can see that the $N_1$ states become mainly populated until $z\approx 0.5$ and that the final $B-L$ asymmetry is much smaller than it is around $z\approx 1$ due to the discussed cancellations. For the final value of $|N_{B-L}|$ the numerics give $1.0\cdot 10^{-13}$, which agrees well with our rough estimations in eq.(\ref{eq final weak wo}).}}
	\end{figure}

After the temperature has dropped below $M_1$, inverse decays and $\Delta L=1$ scatterings will be suppressed and the particles will decay around $H\approx \Gamma _D /2$, which corresponds to $z=K^{-1/2}$. The decaying heavy neutrinos will produce an asymmetry, of the same amount as the inverse decays and scatterings that populated these states. However, the sign will be the opposite and therefore the first term in eq.(\ref{eq b-l approx z=1}) will be exactly canceled, which shows that we had to include the wash-out effects to find a non-zero asymmetry, which then amounts to
\begin{equation}\label{eq final weak wo}
	N_{B-L}(z=\infty)\approx - \varepsilon _1  K^2 \, .
	\end{equation}
A numerically calculated solution can be found in fig.\ref{fig weak wash-out 1 part}, which also illustrates the basic ideas.
	
\subsection{The strong wash-out scenario} \label{sec The strong wash-out scenario} 

In the strong wash-out scenario ($K\gg 1$), the wash-out processes come into equilibrium, and therefore the $B-L$ asymmetry, which is created during the population of $N_1$ states, is washed out before the states decay again. Therefore we do not have a cancellation of the two asymmetries as in the weak wash-out scenario. 

On the other hand, the large $K$ also keeps the right-handed neutrino density close to its equilibrium value, which means that the majority of the right-handed neutrinos will decay while the wash-out processes are still active. This of course reduces the final asymmetry.

Let use some rough approximations, again. After $z\approx 1$ the wash-out processes will get an additional exponential suppression factor, since the right-handed neutrinos now have masses larger than the temperature and need to be produced on-shell in all the wash-out processes considered in this part of the paper. With regard to eq.(\ref{eq approx Gamma W weak-wo}) we now find (cf. \cite{Kolb:1990vq})
\begin{equation} \label{eq wash-out z<1}
	\Gamma _W \approx  \, \left( \frac 12 \cdot \frac 1{8\pi}
	(gg^\dagger)_{11}M_1 \,  
	+ \frac{|g_{\textrm{top}}|^2}{\pi ^3}\,
			 (gg^\dagger)_{11}\, T \right)
	z^{3/2} e^{-z} 
	\, ,
	\end{equation}
for $z \gtrsim 1$ which shows that we can neglect the part due to scatterings at late times.

In a naive approximation the wash-out processes will no longer be in thermal equilibrium and therefore not be able to erase a produced asymmetry when $\Gamma _W \lesssim 2 H$. However, due to the rapid suppression of this term, we can make a better approximation. As a result of the wash-out processes an asymmetry will behave as
\begin{equation}
	N_{B-L}(t)=N_{B-L}(t_0) \exp \left( - \int _{t_0}^t 
		\Gamma _W (t) dt \right)
		=  N_{B-L}(z_0) \exp \left( - \int _{z_0}^z 
			\frac {\Gamma _W(z)}{H(z) z}  dz \right)	\, .
	\end{equation} 
As $\Gamma _W$ is very rapidly suppressed, the exponential term can be neglected as soon as the integrand is smaller than one. Therefore, a more exact condition for the ``time'' $z_*$ at which the wash-out processes become ineffective is
\begin{eqnarray} 
	\frac {\Gamma _W}{z_*} &\approx &  H \\
	\label{eq rate comp strong wo}
	 \Rightarrow  z_*^{5/2} e^{- z_*} &\approx & K^{-1} \, .
	\end{eqnarray}

\begin{figure}
	\begin{center}\includegraphics[%
	  width=0.5\columnwidth,keepaspectratio]
	  {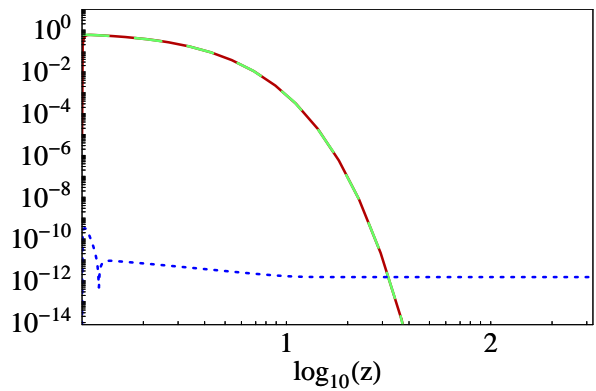}
	  \end{center}
	\caption{\label{fig strong wash-out 1 part}\textit{A typical example for a strong wash-out scenario. The green/dashed  line shows the equilibrium density of the right-handed neutrinos $N_1$ per relativistic photon and almost perfectly matches the red/solid line, which illustrates their actual density as calculated from the Boltzmann equations. The blue/dotted line shows the absolute amount of the $B-L$ density. This graph was plotted with $K=10^{2}$ and all the other parameters as in figure \ref{fig weak wash-out 1 part}. One can see that the $B-L$ asymmetry from the production of $N_1$ states is almost immediately washed-out. For the final value of $|N_{B-L}|$ we find $1.5 \cdot 10^{-12}$, which matches our approximative value in eq.(\ref{eq final strong wo}) roughly to a factor of 3.}}
	\end{figure}

By this time, the equilibrium number density of right-handed neutrinos has decreased and we have 
\begin{equation}
	N(z_*)\approx N^\textrm{eq}(z_*) 
	\approx z_*^{3/2}\exp(-z_*)/2 \approx (2z_*K)^{-1} 
	\approx \frac {0.05}K \, .
	\end{equation}
Therefore our final result for the $B-L$ asymmetry is
\begin{equation}\label{eq final strong wo}
	N_{B-L}(z=\infty) 
	\approx  -\varepsilon _1 \frac {0.05}{K} \, .
	\end{equation}
An example for a numerical solution of the corresponding Boltzmann equations is given in fig. \ref{fig strong wash-out 1 part}, which supports our rough estimations.
	
\subsection{Upper mass bound for $\mathbf{\overline{m}^2}$} 
	\label{sec upper bound mbar}
	
The strong wash-out scenario also places an upper limit on $\overline{m}^2$ \cite{Fukugita:1990gb,Harvey:1990qw,Nelson:1990ir,Buchmuller:1992qc}. Let us illustrate the origin of this bound along the discussion in ref. \cite{Abada:2006fw}, with slightly different numerical factors.

In case of a large $\overline{m}^2$, all the light neutrinos are degenerate. This also implies \cite{Fujii:2002jw}
\begin{equation} \label{eq upper mass consequence}
	m_1 \approx \tilde m_1 \approx m_3 \approx \overline{m}/\sqrt{3} \, 
	\quad \textrm{and} \quad 
	K\approx \frac {\bar m }{\sqrt{3} m_*} \, ,
	\end{equation}
which tells us that we are in the strong wash-out regime. Thus, the freeze out of $\Delta L=2$ scatterings around $z=1$ is crucial to avoid the wash-out of the generated $B-L$ asymmetry. Estimating the rate for these processes to be
\begin{equation} \label{eq eps for upper boun bar m}
	\Gamma _{\Delta L=2} \approx \frac {\overline{m}^2 T^3}
		{ \pi ^3 \langle \phi  \rangle ^4} \, ,
	\end{equation}
the condition $\Gamma _{\Delta L=2}(z\approx 1) \ll H(z\approx 1)$ translates into the upper mass limit 
\begin{equation} \label{eq upper bound M1}
	M \lesssim \left(  \frac {\textrm{eV}}{\overline{m}} \right) ^2
	 5\cdot 10^{10}\textrm{GeV}.
	\end{equation}

With eq.(\ref{eq upper mass consequence}) we can also transform eq.(\ref{eq davidson ibarra more general}) to 
\begin{equation}\label{eq upper bound m DI}
	\varepsilon _1 \lesssim  \frac{3 M_1 \Delta m^2_\textrm{atm}}
	{16 \pi \langle \phi  \rangle ^2
	 \overline{m}} \, .
	\end{equation}
		
Combining this with equations (\ref{eq measured asym}), (\ref{eq dilution factor}), (\ref{eq final strong wo}), and (\ref{eq upper mass consequence}) we find the rough upper mass limit
\mbox{$\overline{m}/\sqrt{3} \lesssim 0.1$eV}. In a more precise analysis refs.\cite{Buchmuller:2003gz,Buchmuller:2004nz} find slightly higher bounds bounds.

\subsection{Lower mass bound for $M_1$}

As is well-known, the assumption that the BAU is due to leptogenesis puts a lower mass bound on the lightest of the heavy neutrinos (if they are hierarchical). Putting the Davidson-Ibarra bound (eq.(\ref{eq davidson ibarra})) together with eq.(\ref{eq dilution factor}) and our estimates for the $B-L$ asymmetry from the Boltzmann equations (eqs.(\ref{eq final weak wo}) and (\ref{eq final strong wo})) we find
\begin{equation}
	\eta _B \lesssim 
	\left\{ \begin{array}{cl}
	
	  10^{-9} K^2 \left(\frac {\displaystyle m_3}
	  {\displaystyle 0.05 \textrm{eV}} \right) 
	   \left(\frac {\displaystyle M_1}
	   {\displaystyle 10^9 \textrm{GeV}} \right) \, , &
	   \textrm{weak wash-out;} \vspace{1ex} \\
	  5 \cdot 10^{-11} K^{-1}
	  \left(\frac {\displaystyle m_3}{\displaystyle 0.05 \textrm{eV}}
	   \right) 
	   \left(\frac {\displaystyle M_1}
	   {\displaystyle 10^9 \textrm{GeV}} \right) \, , &
	   \textrm{strong wash-out.} 
	   \end{array} \right.
	\end{equation}
We see that the region around $K\approx 1$ will give the largest baryon asymmetry. Reference \cite{Buchmuller:2002rq} finds 
\begin{equation} \label{eq stand lept low bound M1}
	M_1 \gtrsim 2 \cdot 10^9 \textrm{GeV}
	\end{equation}
as the lowest bound for the lightest of the singlet neutrino masses in case of zero initial abundance.

\subsection{Complications}

Several effects have been neglected so far, and we now wish to comment on them:

{\bf $\mathbf{\Delta L= 1}$ scatterings:} \\
We have neglected $\Delta L= 1$ scatterings that include gauge bosons. In this paper we adopt the treatment from reference \cite{Buchmuller:2004nz}, assuming that these processes contribute at most in the same way as those scatterings that involve the top quark. Being mainly concerned with orders of magnitude, this should not change our results. (Also, the neglect of the running of the top Yukawa coupling should partially compensate for this.)

{\bf $\mathbf{\Delta L= 2}$ scatterings:}
\label{sec complications}
\\
If these scatterings are in thermal equilibrium at late times, they can wash-out the created asymmetry. Based on dimensional analysis we estimate their rates to be $\Gamma _{\Delta L=2} = \overline{m}^2 T^3/({\pi ^3 \langle \phi  \rangle ^4}) $. The condition that these processes are out of equilibrium around $z\approx 1$ (cf. also \cite{Abada:2006ea}) then translates to
\begin{equation} \label{eq delta Leq2 up bound M1}
	M \lesssim \frac {17 \, \pi ^3 \langle \phi  \rangle ^4}
		{\overline{m}^2 \MPl} 
	\approx 2\cdot 10^{13}
	\textrm{GeV.}
	\end{equation}
In the weak wash-out scenario these processes can still be in effect around $z=1$, but not at the time of decay $z\approx (2 /K)^{1/2}$. The mass bound then changes to
\begin{equation}
	M \lesssim \sqrt{\frac 2K} 
	\frac {17 \, \pi ^3 \langle \phi  \rangle ^4}{\overline{m}^2 \MPl} 
	\approx \sqrt{\frac 2K}\cdot 2\cdot 10^{13}
	\textrm{GeV.}
	\end{equation}

{\bf Quantum Effects:} \\
Quantum effects can also play a larger role in leptogenesis scenarios than indicated by our treatment.

Examples are 
the running of coupling constants  \cite{Antusch:2003kp} that can affect the final asymmetry by an order one factor and thermal quantum effects within the treatment of Boltzmann equations \cite{Giudice:2003jh} that can also have an impact on the final asymmetry.

Moreover, the semi-classical treatment with Boltzmann equations is of limited validity compared to the full quantum dynamics described by Kadanoff-Baym equations (see e.g. \cite{Lindner:2005kv}). 
In spite of recent progress on the subject (e.g. \cite{Buchmuller:2000nd,DeSimone:2007rw}) a full quantum mechanical treatment of leptogenesis in terms of Kadanoff-Baym equation has not been performed, yet. 

{\bf Spectator Processes and Flavor effects:} \\
The standard treatment for leptogenesis ignores the possible change of chemical potentials of particles other than the SM leptons. However, this possible time dependence through spectator processes in equilibrium can contribute an order one suppression \cite{Buchmuller:2001sr}.

Lately, much attention has also been given to flavor effects in leptogenesis scenarios, e.g. refs. \cite{Barbieri:1999ma,Pilaftsis:2003gt,Abada:2006fw,Nardi:2006fx,Abada:2006ea}. The important fact, which had been neglected before, is that at low reheating temperatures ($T_r \lesssim 10^{12}$GeV), the Yukawa couplings for the third (and sometimes second) generation leptons are in thermal equilibrium at the time of leptogenesis. This allows us to distinguish lepton flavors at that point, and hence one has to use a system of coupled Boltzmann equations. Naively, this yields an additional factor of 3 for the final asymmetry in the weak- and the strong wash-out scenario, while larger discrepancies can also be achieved. However, the overall lower bound for $M_1$ cannot be reduced by flavor effects \cite{Blanchet:2006be}.
For simplicity, we will only be dealing with  the one-flavor approximation in this paper.


\section{Leptogenesis With Many Heavy Neutrinos}

The purpose of this chapter is to examine the effect of additional heavy neutrino states on standard thermal leptogenesis especially in the context of the necessary reheating temperature. The most important result in this context is that this bound can be significantly lowered without any resonant enhancement of the CP asymmetries.

To be more specific, we consider a setup with $n_N$ right-handed neutrinos, of which $n\eff$ have masses that are low enough such that they will reach significant abundances during the leptogenesis process (i.e. they have masses not much higher than the reheating temperature). 
In this case, the Lagrangian of this part of the particle spectrum is still given by eq.(\ref{eq singlet addition to L}), with the difference that the index $j$ is now running from 1 to $n_N$, and $g$ is a $n_N \times 3$-dimensional matrix.

The see-saw formula is also not significantly altered, since the diagram in fig.\ref{fig see-saw} still gives the effective mass of the light neutrinos to leading order (cf. also \cite{Dienes:1998sb})
\begin{equation}
	m_{\nu,ik} = g^T_{ij}M_j^{-1}g_{jk} \langle \phi \rangle ^2 \, ,
	\end{equation}
again with the difference that $g$ is now a $n_N \times 3$-dimensional matrix.

We see that the Yukawa couplings tend to be smaller compared to the standard case with similar masses for the singlets, if there are no cancellations between the coupling constants or only few dominating couplings. This already implies that the weak wash-out regime can be natural in scenarios with many singlets (cf. ref. \cite{Buchmuller:2007zd} for a related argument).

The Boltzmann equations from the previous section (eqs.(\ref{eq boltzmann standard N}) and (\ref{eq boltzmann standard B-L})) are easily generalized to
\begin{equation} \label{eq boltzmann many N}
	\frac {dN_{N_n}}{dz} = - \frac {1}{Hz} 
	(\Gamma _{D,n} + \Gamma _{\Delta L =1,n}) \, 
	(N_{N_n}-N^\textrm{eq}_{N_n}) \, ,
	\end{equation}
with $1\le n \le n\eff$. One can calculate $\Gamma _{D,n}$ and $\Gamma _{\Delta L =1,n}$ as in the standard case, with the corresponding interchange of masses and coupling constants (cf. appendix \ref{sec app reac rates}). 

The second equation now reads
\begin{equation} \label{eq boltzmann many B-L}
	\frac {dN_{B-L}}{dz} = 
	- \frac {1}{Hz} \sum _{n=1}^{n_\textrm{eff}} \left[
	 \varepsilon _n (\Gamma _{D,n} + \Gamma _{\Delta L =1,n}) \, 
	(N_{N_n}-N^\textrm{eq}_{N_n})\,
	+ \Gamma _{W,n} N_{B-L} \right] \, ,
	\end{equation}
where $\Gamma _{W,n}$ can also be calculated as in the previous section, again with the corresponding interchange of masses and coupling constants. $\varepsilon _n$ is discussed in detail in the following section and Appendix \ref{sec app DI}.
	
\subsection{The Davidson-Ibarra bound with many heavy neutrinos}
	\label{sec davidson ibarra bound}

Let us first examine the CP asymmetry for the decay process of one of the right-handed neutrinos. As in the three generation case, we can draw wave-function and vertex corrections as in fig. \ref{fig lept graphs}.
However, once more with the difference that we now have $n_N$ of each sort of the loop diagrams instead of 3, as we did in the standard case.%
\footnote{In principle, the large number of possible diagrams can invalidate this perturbative approach. However, for Yukawa couplings that yield the light neutrino masses without delicate cancellations this is typically not the case.}
To calculate the CP asymmetry we can simply add up the various contributions and find (as in eq.(\ref{eq CP standard}))
\begin{eqnarray} 
	\varepsilon _i &=& \frac {\Gamma(N_i \rightarrow  \ell_L + \Phi)-
	  	\Gamma(N_i \rightarrow  \bar \ell_L + \Phi ^*)}
		{\Gamma(N_i \rightarrow  \ell_L + \Phi)+
	  	\Gamma(N_i \rightarrow  \bar \ell_L + \Phi ^*)} \\
		&= &- \frac 3{16 \pi}
		\sum _k \frac {\textrm{Im}\left[ (g g^\dagger)^2_{ik} \right]}
		{(g g^\dagger)_{ii}} \, \displaystyle \frac {M_i}{M_k} 
		+{\cal O}\left( \frac {M_i^2}{M_k^2} \right)          
		\label{eq CP asym many nus 1} \\
		&= &- \frac 3{16 \pi} \frac {M_i}{\langle \phi \rangle ^2}
		\frac 1{(g g^\dagger)_{ii}} 
		{\textrm{Im}\left[ (g  m_{\nu}^\dagger g^T )_{ii} \right]}
		+{\cal O}\left( \frac {M_i^2}{M_k^2} \right) 
		\label{eq CP asym many Ns}
	\end{eqnarray}
with $k$ running from 1 to $n_N$, now. 

Of course this result is only appropriate if the higher-order corrections in eq.(\ref{eq CP asym many nus 1}) (cf. ref. \cite{Hambye:2003rt}) and resonance effects can be neglected, or if $\textrm{Im}\left[ (g g^\dagger)^2_{ik} \right]$ vanishes. For simplicity we restrict ourselves to this case in the following. 

As we show in Appendix \ref{sec app DI}, eq.(\ref{eq CP asym many Ns}) leads to the relation%
\footnote{The author is very grateful to Yosef Nir for pointing out to him that the Davidson-Ibarra bound does lose its validity in the case of many singlets and three light neutrinos, due to the fact that the matrix $R$ within the Casas-Ibarra parametrization \cite{Casas:2001sr} of the Yukawa couplings is not orthogonal anymore in the present case.}
\begin{equation} \label{eq DI many Ns v2}
	|\varepsilon _1| \lesssim 
	 \frac 3{16\pi} \frac{M_1 m_3}{\langle \phi \rangle ^2}\approx 
	  10^{-7}\left(\frac {m_3}{0.05 \textrm{eV}} \right) 
	   \left(\frac {M_1}{10^9 \textrm{GeV}} \right) \, 
	   \end{equation}
which matches eq.(\ref{eq davidson ibarra}). Yet, in the latter case the CP-asymmetry is maximized for $m_1=0$ and $m_3$ is therefore fixed around 0.05eV. In the presence of many singlets $m_3$ is in general a free parameter in eq.(\ref{eq DI many Ns v2}), and therefore the standard Davidson-Ibarra bound is considerably weakened.

However, as we shall see in section \ref{sec An Extra-Dimensional Example}, the Davidson-Ibarra bound in eq.(5) still applies in many cases.
 This is due to the fact that the number of free parameters for the Yukawa couplings can be reduced by symmetries. In particular, the condition of maximal CP asymmetries for all decaying particles can motivate such effects. Thus, we will not consider possible CP asymmetries beyond the Davidson-Ibarra bound in eq.(\ref{eq davidson ibarra}) in the following analysis. Nevertheless, it should be kept in mind that there might be further possibilities to increase the final baryon asymmetry due to the arguments given in this subsection.

At this point it is also important to note that the analysis in ref. \cite{Abada:2006ea} that determines the CP-asymmetry in $\Delta L=1$ scatterings is independent of the number of heavy neutrinos. Thus, the CP-asymmetries of these processes match the ones of the corresponding decays as in the three neutrino case.

\subsection{The strong wash-out scenario}

Let us first see, how the strong wash-out scenario may be changed by the presence of many heavy neutrinos. First, we note that if we have many neutrinos with $K_i\gg 1$, it is enough to observe the lightest ones, i.e. only those particles that are still abundant in unsuppressed numbers, when the wash-out processes due to the lightest one of the singlets freezes out, since all asymmetries produced by heavier particles will have been erased by these processes.\footnote{
This is only true within the one-flavor approximation considered here. As shown in ref.\cite{Engelhard:2006yg} things can change once flavor is considered.} %
Therefore, we consider the case of $n\eff$ quasi-degenerate neutrinos.

The addition of extra singlet states has two competing effects on the scenario. The first one is the fact that the number of decaying neutrinos is increased, which would increase the total amount of the produced $B-L$ asymmetry if it was the only change. However, the wash-out rate is also increased, which means the time after which a produced asymmetry is not washed-out anymore is later. This  effectively reduces the abundance of decaying neutrinos.

Let us make some approximations in the manner of section \ref{sec The strong wash-out scenario}. The averaged wash-out rate is then (compare eq.(\ref{eq wash-out z<1}))
\begin{equation} \label{eq wash-out z<1 manny nu}
	\Gamma _W \approx  \sum _{i=1}^{n\eff} 
	 \left( \frac 12 \cdot \frac 1{8\pi}
	(gg^\dagger)_{ii}M_i \,  
	+ \frac{|g_{\textrm{top}}|^2}{\pi ^3}\,
			 (gg^\dagger)_{ii}\, T \right)
	z_i^{3/2} e^{-z_i} 
	\, ,
	\end{equation}
with $z_i \equiv M/T_i$.

\begin{figure}
	\begin{center}\includegraphics[%
	  width=0.5\columnwidth,keepaspectratio]
	  {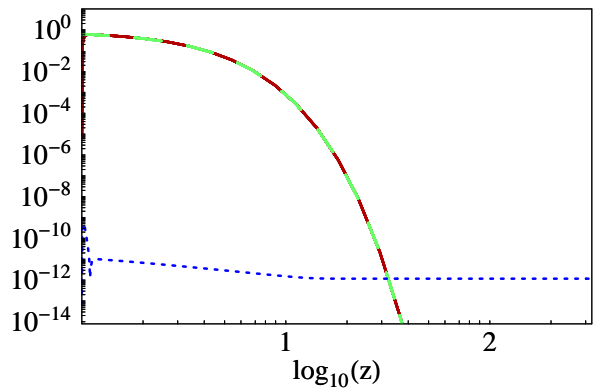}
	  \end{center}
	\caption{\label{fig strong wash-out 10 part deg}\textit{The numerical solutions for a scenario with 10 decaying singlets that is in all other respects identical to the setup in in \mbox{figure \ref{fig strong wash-out 1 part}}. The green/dashed line is the equilibrium abundance per comoving volume of one of the particle species, the red/solid line is its actual abundance as given by the Boltzmann equations, and the blue/dotted line is the absolute amount of the corresponding density of $B-L$. For the final value of the last quantity we find $1.2\cdot 10^{-12}$ and therefore a slightly smaller value than in the corresponding one-particle scenario, which agrees with our considerations.}}
	\end{figure}

This means the freeze-out ``time'' of wash-out processes $z_*$ is now defined by the equation
\begin{equation} \label{eq rate comp strong wo manny nu}
	z_*^{5/2} e^{- z_*} \approx  (\sum K_i)^{-1} \, .
	\end{equation}
The number density of each species of right handed neutrinos at this point is
\begin{equation}\label{eq eff heavy nu number many nu}
	N_i(z_*)= \frac 1{2 z_*} \frac 1{\sum K_i} 
	\approx \frac {0.05}{\sum K_i} \, ,
	\end{equation}
which leads to an approximate final $B-L$ asymmetry of
\begin{equation}\label{eq final strong wo manny nu}
	N_{B-L}(z=\infty) 
	\approx - 0.05 \frac {\sum \varepsilon _i}{\sum K_i} \, .
	\end{equation}
If all $n\eff$ singlets are to take part in the process, the final asymmetry will be the largest, when $K_i=\textrm{Min}(\{K_j, 1\leq j \leq n\eff\})$ and $\varepsilon_i=\textrm{Max}(\{\varepsilon_j, 1\leq j \leq n\eff\})$ for all $i$, which makes the different particle states identical within our treatment. However,
we can also see that in this case the final baryon asymmetry does not change  compared to the one particle case within our approximation, since we get a factor of $n\eff$ in the numerator and the denominator of eq.(\ref{eq final strong wo manny nu}). A closer look at eq.(\ref{eq eff heavy nu number many nu}) shows that the asymmetry will even be slightly reduced compared to the one particle case due to the factor $z_*^{-1}$, which tends to smaller values. A numerical example that verifies our considerations is shown in figure \ref{fig strong wash-out 10 part deg}, where we consider the same scenario as in section \ref{sec The strong wash-out scenario} with 10 identical neutrinos. 

\subsection{The weak wash-out scenario in a radiation dominated universe}

As we have seen, in the strong wash-out regime the presence of many heavy neutrinos cannot lower the mass bound for heavy neutrinos compared to the standard case . This is due to the competing effects of the singlets on particle number and the wash-out rate. In the weak wash-out regime, this is different, as a result of the fact that an increase of the wash-out rate enlarges the final asymmetry. Therefore the increase of the number of decaying neutrinos and the increase of the wash-out rate, will not partially annihilate each other's effect. Instead both of them will amplify the final asymmetry.

Let us make approximations similar to those in section \ref{sec The weak wash-out scenario}. 
In concordance with equation (\ref{eq abund N weak}) the abundance of each neutrino species with mass around the reheating temperature or lower is given by
\begin{equation} \label{eq no dens many nu}
	N_{N_i}(z\approx 1) \approx  
		\frac {(\Gamma _{D,i} + \Gamma _{\Delta L =1,i})N^\textrm{eq}_{N_i}}
		{H(z=1)} \approx  K_i \, .
	\end{equation}	
The wash-out rate is the sum over the rates of all possible inverse decays and $\Delta L=1$-scatterings for which we can write
\begin{equation} \label{eq approx Gamma W weak-wo many nu}
	\Gamma _W \approx \sum _{i=1}^{n\eff} 
	\frac 12 \Gamma _{ID,i} + \frac 23 \Gamma _ {\Delta L=1,i}
	\approx \sum _{i=1}^{n\eff} \Gamma _{D,i}
	\, ,
	\end{equation}
in case of $T\gtrsim M_i$.

In the same manner in which we reached eq.(\ref{eq b-l approx z=1}), we now find
\begin{eqnarray}
	N_{B-L}(z=1) &\approx & 
		 \sum _{i=1}^{n\eff} \varepsilon _i N_{N_i}(z\approx 1)
	\exp \left(- \frac {\Gamma _W}{H(z=1)} \right) \\
	&\approx &  \left( \sum \varepsilon _i   K_i \right) 
		\exp \left(- \sum K _i\right) \, ,
	\end{eqnarray}
with the difference that now we are generally not allowed to make the linear approximation of the exponential function around zero. This is due to the fact the sum over the various $K_i$ can be bigger than one, even though each $K_i$ is much smaller.  

The late-time decay of the particles yields an asymmetry of $\sum  \varepsilon _i N_{N_i}(z\approx 1)$. Thus, combining inverse decay and decay periods, we are lead to the final asymmetry
\begin{equation}
	N_{B-L}(z\rightarrow \infty) \approx  
		 \left( \sum \varepsilon _i   K_i \right) 
		\left[\exp \left(- \sum K _i\right) -1 \right] \, .
	\end{equation}
In the case $\sum K _i \ll 1$, this leads to
\begin{equation} \label{eq bar as approx weak semi-many N}
	N_{B-L}(z\rightarrow \infty) \approx  
		 - \left( \sum \varepsilon _i   K_i \right) 
		\left(\sum K _i\right)  \, ,
	\end{equation}
while we can approximate
\begin{equation} \label{eq bar as approx weak many N}
	N_{B-L}(z\rightarrow \infty) \approx  
		 - \sum \varepsilon _i   K_i 
		 \, ,
	\end{equation}
in the case $\sum K _i > 1$.%
\footnote{In the latter case, the phrase ``weak wash-out'' might be confusing, since the wash-out processes are in equilibrium around $z \approx 1$. However, around the decay time of the singlets, these processes will typically be frozen out. Therefore, we still refer to this scenario as a ``weak wash-out scenario''.}

\begin{figure}
	\begin{center}\includegraphics[%
	  width=0.45\columnwidth,keepaspectratio]
	  {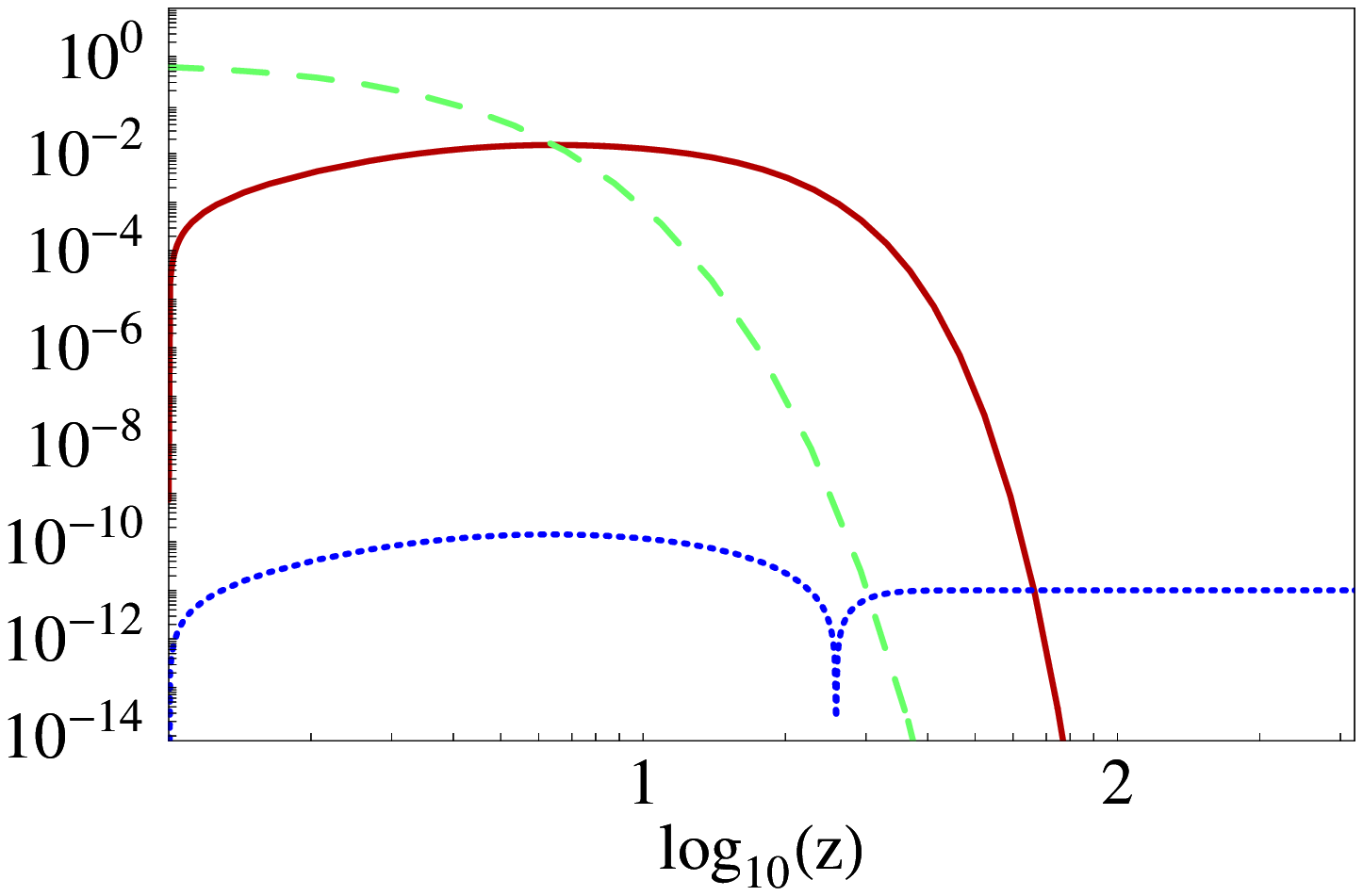}
	  \includegraphics[%
	  width=0.45\columnwidth,keepaspectratio]
	  {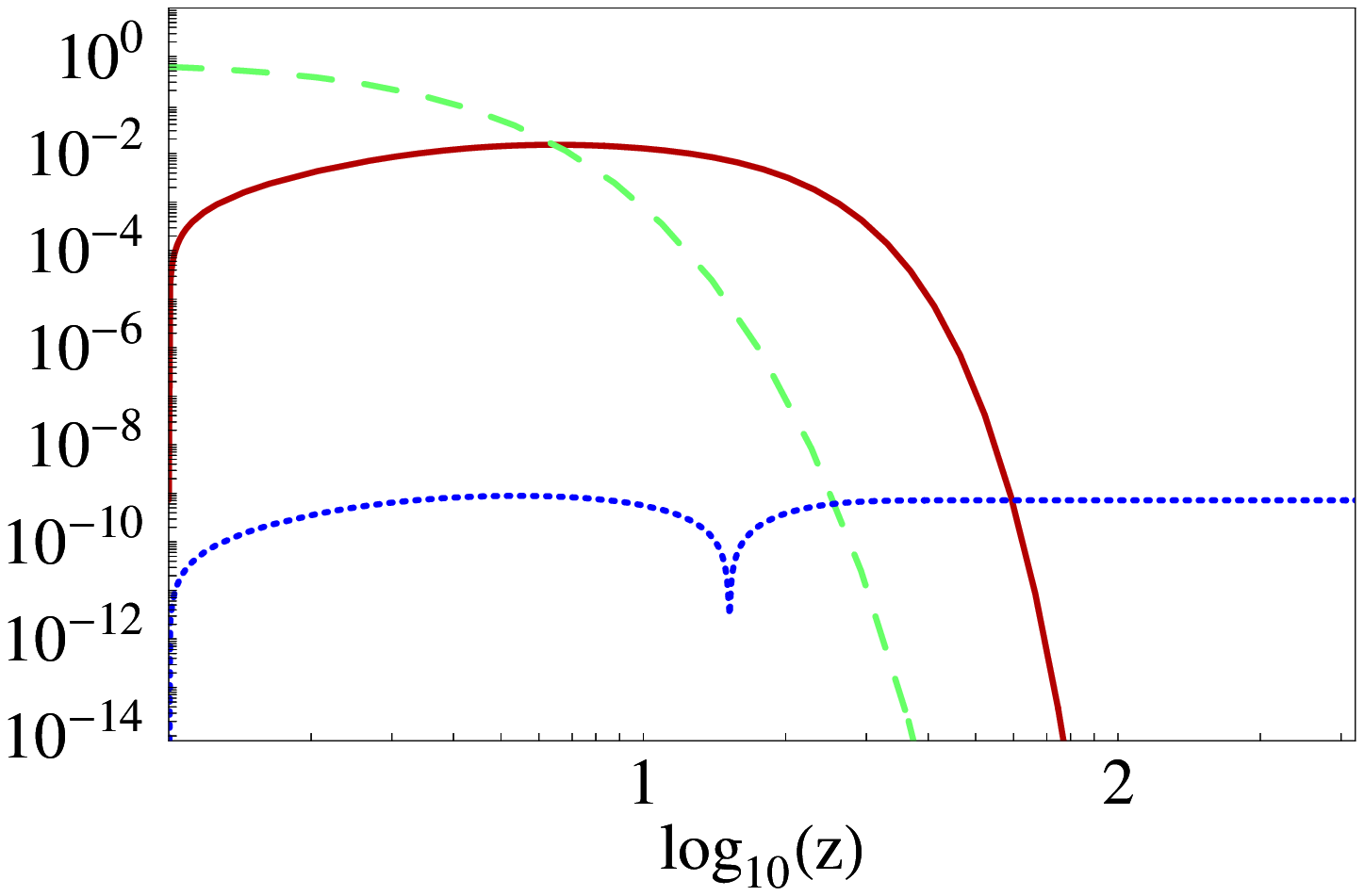}
	  
	  \end{center}
	\caption{\label{fig weak wash-out 10 part deg}\textit{Here, we give two examples for leptogenesis with many singlets in the weak wash-out regime, where we used the same parameter values as in the setup considered in figure \ref{fig weak wash-out 1 part}. The red/solid line shows the abundance of particles of one singlet species in a comoving volume, whereas the green/dashed line illustrates the abundance of the corresponding equilibrium value. The blue/dotted line still represents the absolute amount of the corresponding $B-L$ asymmetry. For the first picture we used $n\eff=10$ and the absolute amount of the final asymmetry is found to be $|N_{B-L}(t\rightarrow \infty)|=1.0 \cdot 10^{-11}$, which is by a factor of $100$ larger than the corresponding one particle scenario, as suggested by our approximative approach (cf. eq.(\ref{eq bar as approx weak semi-many N})). In the second graph $n\eff=100$ was used. Now, the wash-out processes are in equilibrium and the final asymmetry is found to be $|N_{B-L}(t\rightarrow \infty)|=7.2 \cdot 10^{-10}$, which also roughly matches our estimation for this case (cf. eq.(\ref{eq bar as approx weak many N})).}}
	\end{figure}

Hence, it is evident that the presence of many right-handed neutrinos can easily increase the final $B-L$ asymmetry in the weak wash-out regime. In the case of similar masses and couplings, the asymmetry increases proportional to $n\eff^2$ for smaller $n\eff$ and has a linear dependence on $n\eff$ when $\sum K _i > 1$. This is also illustrated by the numerical solutions shown in figure \ref{fig weak wash-out 10 part deg}.

Of course these approximations are only valid, as long as $n\eff$ does not get so big that the right-handed neutrinos get to dominate the energy density of the universe. In this case, the energy density of the singlets would become comparable to the background and they would produce a significant amount of entropy during their decay, hereby diluting the asymmetry. This possibility will be considered in the next section

To get a feeling up to which point our estimations from this chapter are still valid, let us again assume that all of the decaying singlets have approximately identical masses and couplings ($K_i \equiv K \ll 1$). If we postulate that the corresponding energy density never comes to dominate the universe, the strongest bound on the number of states will arise at the time of their decay. 
In this case, we can assume to be in a background dominated universe at early times, which implies that we can still use eq.(\ref{eq no dens many nu}) to determine the abundance of the various singlets. 

Once the states are populated, they will decay around the time 
\begin{equation} \label{eq time od decay}
	\Gamma ^{-1} = K^{-1} H^{-1}(z=1),
	\end{equation}
where t=0 corresponds to $z=1$.

By then, the radiation density will have decreased to
\begin{eqnarray}
	\rho _\gamma &=& \frac {\pi ^2}{30} \, g_* \, T_0^4 \left( 
	\frac {a(z=1)}{a(t=\Gamma ^{-1})}
		\right) ^4 \\
		&\approx & 35 \, M^4 \, (2 K^{-1})^{-\frac 42} \\ 
		&\approx & 9 \, M^4 K^2 \, ,
	\end{eqnarray}
where $a$ is a comoving length scale.

The energy stored in the right-handed neutrinos will also have decreased
\begin{eqnarray}
	\rho _N &\approx& n\eff \, K \left( 2 \, \frac {\xi (3)}{\pi ^2} 
		 T_0^3 \right) M 
		\left( 2K^{-1}
		\right) ^{-\frac 32} \\
		&\approx &  0.1 \cdot n\eff \,  M^4 K^{\frac 52} \, ,
	\end{eqnarray}
with $\xi (3)\approx 1.2$ \cite{Kolb:1990vq}.

\begin{figure}
	\begin{center}\includegraphics[%
	  width=0.45\columnwidth,keepaspectratio]
	  {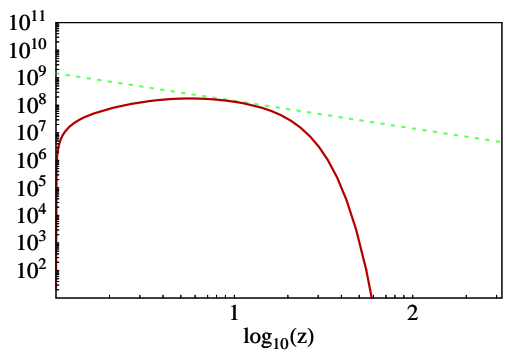}
	  
	  \end{center}
	\caption{\label{fig weak wash-out 900 part energy ratio}\textit{Here, we show the energies stored in a comoving volume containing one photon, as calculated from the numerical solutions  of eqs.(\ref{eq boltzmann many N}) and (\ref{eq boltzmann many B-L}) in a weak-wash out scenario ($K=0.01$) with 900 different singlet states and parameter values as in the corresponding figures before. The energy due to radiation is illustrated by the green/dashed line, whereas the energy due to the singlets corresponds to the red/solid line. We can see that the energy densities become comparable around the time of decay, as suggested by eq.(\ref{eq upper bound n}).}}
	\end{figure}

The condition that $\rho _N \lesssim \rho _\gamma$ at the time of decay, then translates into the relation
\begin{equation} \label{eq upper bound n}
	n\eff \lesssim \frac {90}{ \sqrt {K}} \, .
	\end{equation}
Also, our numerical analysis in figure \ref{fig weak wash-out 900 part energy ratio} with 900 identical particles and $K=0.01$ shows that the energy densities tend to become comparable around the time of decay if eq.(\ref{eq upper bound n}) is fulfilled.

\subsection{Weak and strong wash-out:}

Before we consider the case of a singlet-dominated universe, let us stick to the background dominated universe a little longer and consider the case, where parts of our set of thermally excited right-handed neutrinos have $K_i\ll 1$ and others $K_i\gg 1$. We will call the corresponding numbers of particle species $n\weak$ and $n\strong $ and further assume that all of these particles are within the same mass range. With the considerations from the previous pages, it is easy to guess, what we can expect from this scenario. Namely, while the presence of the weakly coupled particles will not have an effect on the final asymmetry produced by the strongly coupled ones, this cannot be said for the opposite case. The presence of one strongly coupled right-handed neutrino is enough to wash-out the asymmetry produced during the population of all the different states, including the weakly coupled ones. This implies that the asymmetry produced by the decays of the weakly coupled singlets is not canceled by an asymmetry of the same size but opposite sign, as it was the case in some of the previously considered scenarios. Hence, the final asymmetry in a scenario with one additional strongly coupled singlet can be substantially increased compared to a scenario with only weakly coupled particles of the same number. 

\begin{figure}
	\begin{center}\includegraphics[%
	  width=0.5\columnwidth,keepaspectratio]
	  {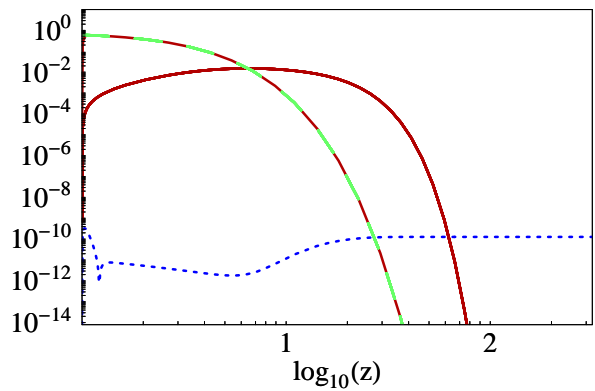}
	  \end{center}
	\caption{\label{fig weak and strong wash-out}\textit{This is an example for a leptogenesis scenario with one strongly coupled right-handed neutrino ($K=10^2$) and ten weakly coupled ones ($K=10^{-2}$). All the other parameter values were chosen as in the examples before. The green/dashed line shows the thermal equilibrium abundance of each of the species in a comoving volume containing one photon. The corresponding actual abundances are illustrated by the red/solid lines, whereas the absolute amount of the $B-L$ asymmetry in the same volume is shown by the blue/dotted line and takes the final value $1.3\cdot 10^{-10}$. We see that the strongly coupled singlets closely follow the equilibrium value and wash out the $B-L$ asymmetry produced during the population of the singlet states. At a later point the weakly coupled particles decay and the main part of the final asymmetry is produced.}}
	\end{figure}

Combining our estimates from the previous sections, we can approximate the final asymmetry by
\begin{equation} \label{eq final asym weak and astrong}
	N_{B-L}(z=\infty) 
	\approx  - 0.05 \frac {\sum _{i=1}^{n\strong} \varepsilon _i}
		{\sum _{i=1}^{n\strong}K_i}
	-  \sum _{i=1}^{n\weak} \varepsilon _i   K_i  \, ,
	\end{equation}
where the sums in the two terms run over different particles, of course.
In figure \ref{fig weak and strong wash-out} we give a numerical example for one strongly coupled and ten weakly coupled particles, which nicely agrees with this estimate.

\subsection{The weak wash-out scenario in a matter dominated universe}

In equations (\ref{eq bar as approx weak many N}) and (\ref{eq final asym weak and astrong}), we have seen that we can increase the final baryon asymmetry of many leptogenesis scenarios by increasing the number of right-handed neutrinos. However, these estimations correspond to solutions of equations (\ref{eq boltzmann many N}) and (\ref{eq boltzmann many B-L}) and are therefore only valid in a universe which is dominated by background radiation and where the energy loss of this background due to the production of the singlet states can be neglected.

However, if we have so many weakly interacting particles that the energy stored in this sector becomes comparable to the radiation energy of the background, their production rate starts to decrease even if they are far below their equilibrium value, since there is less and less energy left to populate these states. Additionally, these particles are massive and weakly coupled, which implies that the energy density of the radiative background will decrease faster than the energy density of the singlets. Thus, if the singlets live long enough they will come to dominate the universe prior to their decay.

In these cases, the dynamics of the system are no longer governed by equations (\ref{eq boltzmann many N}) and (\ref{eq boltzmann many B-L}). Instead, we use the more complex system of equations
\begin{eqnarray} \label{eq boltzmann many N over t}
	\frac {dN_{N_n}}{dt'} &=& - \frac {1}{H_0} 
	(\Gamma _{D,n} + \Gamma _{\Delta L =1,n}) \, 
	(N_{N_n}-N^\textrm{eq}_{N_n}) \, , \\
	 \frac {dN_{B-L}}{dt'} &=& 
	- \frac {1}{H_0} \sum _{n=1}^{n_\textrm{eff}} \left[
	 \varepsilon _n (\Gamma _{D,n} + \Gamma _{\Delta L =1,n}) \, 
	(N_{N_n}-N^\textrm{eq}_{N_n}) \,
	+ \Gamma _{W,n} N_{B-L} \right] \, , \\
	\frac {d(\rho_\gamma a^3)}{dt'} &=&
	- (\rho_\gamma a^3) \frac {H(t')}{H_0}- \sum \frac {d( N_{N_n})}{dt'}
		\langle E_{n} \rangle \, , \\
	\frac {da}{dt'} &=& \frac {H(t')}{H_0} a(t') \, , 
	\end{eqnarray}
where $t'\equiv t H_0$ is the time measured in units of $H_0\equiv 1.66 g_*^{1/2} M_1^2/ \MPl$. $a$ is a comoving length scale with the volume $a^3$ containing one photon at early times (of course, the number of photons in this volume changes with time due to the production and decay of the singlets). $N_{N_n}$ is the abundance of particles of the species $N_n$ in a comoving volume of size $a^3$, and $N_{B-L}$ is the corresponding value for the $B-L$ asymmetry.

Further we have the relations
\begin{eqnarray}
	H(t') &=& \sqrt{\frac {8 \pi}3} \frac 1\MPl 
		\sqrt{ \frac {\pi ^2}{30} g_* T^4 + \sum \frac {N_{N_n}}{R^3}
			\langle E_{n} \rangle} \, ,\\
	\langle E_{n} \rangle &=& 3 \, T + M_n 
		\frac {K_1(M_n/T)}{K_2(M_n/T)} \, , \\
	T &=&  \left( \frac {30}{\pi ^2} g_*^{-1} 
		\rho_\gamma \right) ^{\frac 14} \, , 
		\label{eq T in scen over time}
		\end{eqnarray}
where $\langle E_{n} \rangle $ is the mean energy \cite{Kolb:1979qa} of the  Boltzmann distributed species $N_n$.%
\footnote{Since the singlets are only weakly coupled to the SM lepton doublets and have no further interaction, the Boltzmann distribution can only be considered as an approximation.}

However, it seems to make limited sense to consider arbitrarily large numbers of active singlet states  $n\eff $, due to the fact that once the energy density stored in the right-handed neutrinos becomes comparable to the energy density of the radiative background, there will not be enough energy left to significantly increase the number density of the right-handed neutrinos. Therefore, an increase of $n\eff$ beyond a certain limit does not imply a significant increase of populated singlet states. On the other hand, for very large values of $n\eff$ the radiation energy might be transferred to the singlet sector so quickly that there will not be enough time for the wash-out processes to prevent the cancellation of the asymmetries during creation and decay of the singlets. 

Thus, we restrict our considerations to scenarios, where the number density of each singlet state per photon is still given by eq.(\ref{eq no dens many nu}), which can be quantified by the condition $\rho _{N} (T\approx M_1) \lesssim \rho _\gamma (T\approx M_1)$, which implies
\begin{equation} \label{eq absolute upper bound neff}
	n\eff K M_1^4 \lesssim  \frac {\pi ^2}{30} g_* M_1^4 \, ,
	\end{equation}
where we additionally assumed that all significantly abundant singlet states have similar masses around $M_1$.

In this class of scenarios, the particle number density corresponding to each state at the respective time of decay will be
\begin{equation}
	n_N \left( \frac {a(z=1)}{a(t=\Gamma ^{-1})} \right)^3 
	\approx n\eff K \left( 2 \, \frac {\xi (3)}{\pi ^2} M_1^3 \right)
		\left( \frac 32 K^{-1} \right) ^{-2}
	\approx 0.1 \cdot n\eff  K^3 M_1^3 \, ,
	\end{equation}
where we now need to determine the absolute number density, since the photon number in our original comoving volume will not be constant anymore.

\begin{figure}
	\begin{center}\includegraphics[%
	  width=0.45\columnwidth,keepaspectratio]
	  {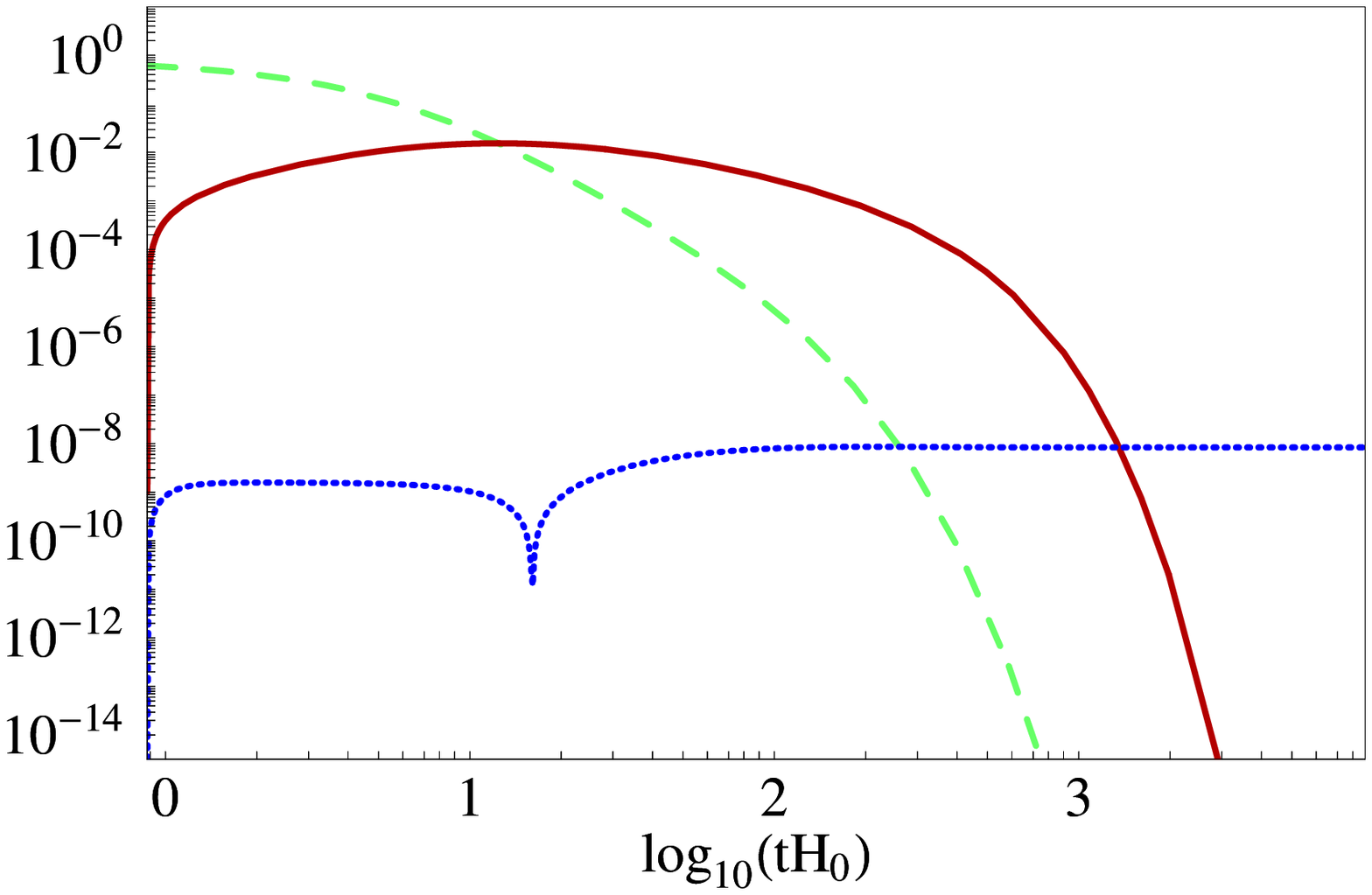}
	  \includegraphics[%
	  width=0.45\columnwidth,keepaspectratio]
	  {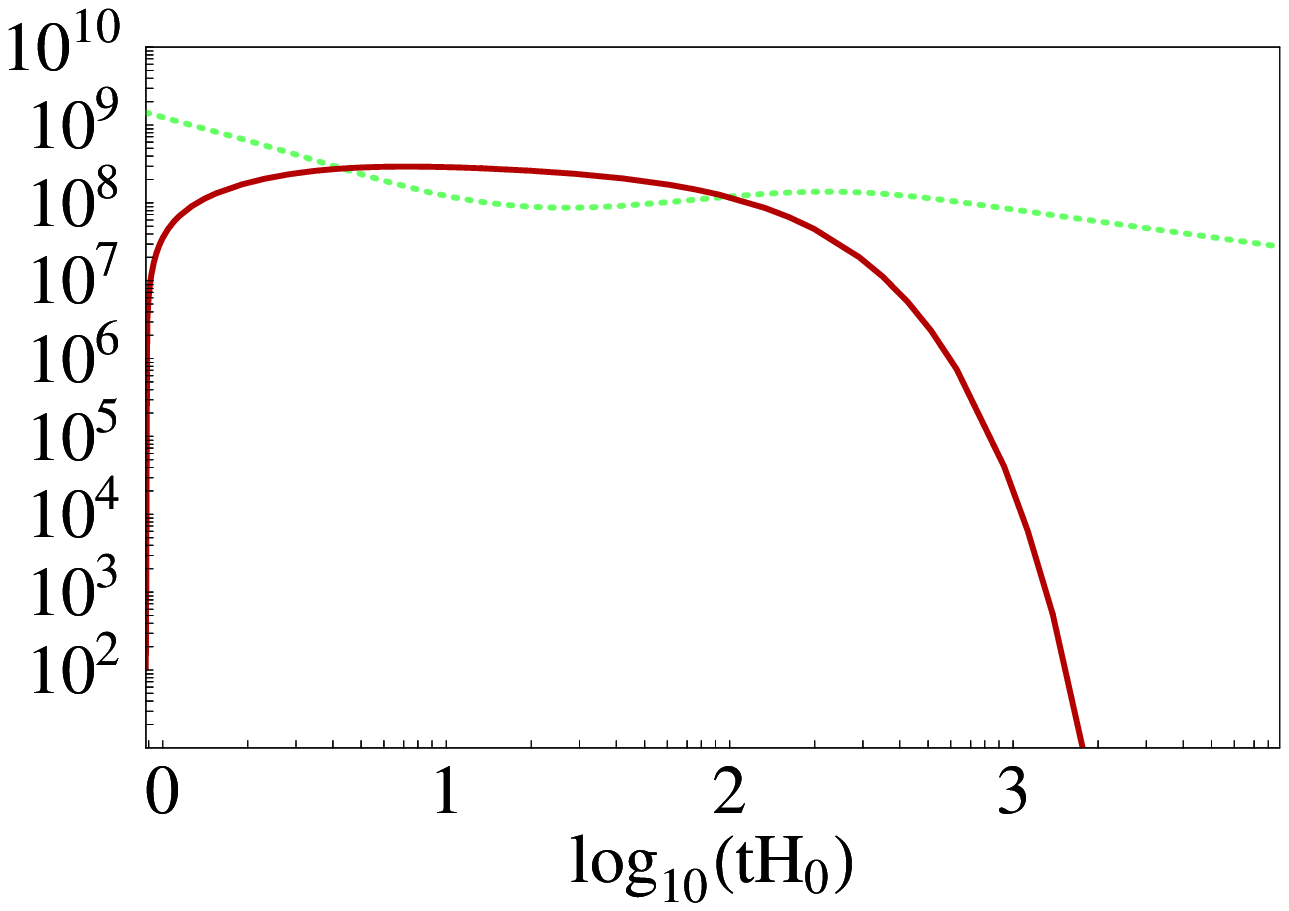}
	  \end{center}
	\caption{\label{fig weak wash-out 3000 part deg over time}\textit{An example for a leptogenesis scenario in a singlet dominated universe. For these two graphs, the parameters $K=10^{-2}$ and $n\eff=3000$ were used, with all other parameters as in the previously considered cases. The left-hand side shows the equilibrium abundance of the singlets (green/dashed) per photon, their actual abundance (red/solid) and the absolute amount of the $B-L$ asymmetry (blue/dotted) per photon which is found to be $8.4 \cdot 10 ^{-9}$ at the end of the calculation, which roughly agrees with our estimation in eq.(\ref{eq weak over time final B-L}). The time $t$ was chosen such that $t=0$ corresponds to the singular point in the previously radiation dominated universe. The graph on the right-hand side, shows the radiation energy in a comoving volume that contains one photon at early times (green/dotted) and the corresponding energy stored in the singlets (red/solid) in GeV. We can see that the singlets get to dominate the universe before their decay.}}
	\end{figure}

From this number density, we can get the energy density ($\rho _N \approx M_1 n_N$), the approximate entropy density after the decay ($s \approx 4 g_*^{1/4} \rho ^{3/4}/3$ \cite{Kolb:1990vq}), and the number density of photons

\begin{equation}
	n_\gamma \approx  2 \, \frac {\xi (3)}{\pi ^2}
		 \frac {45}{2 \pi ^2 g_*}s \approx 
	0.4 \cdot 10^{-2}\, n^{\frac 34}\eff \,K^{\frac 94}\, M_1^3 \, .	 
	\end{equation}
Thus, our final value for the $B-L$ asymmetry per photon (at early times) is
\begin{equation} \label{eq weak over time final B-L}
	\frac {n_{B-L}}{n_\gamma} \approx 
		- \varepsilon _1\,  25 \, n^{\frac 14}\eff
			\, K^{\frac 34} \, .
	\end{equation}
A numerical solution using eqs.(\ref{eq boltzmann many N over t})-(\ref{eq T in scen over time}) for this scenario is shown in fig. \ref{fig weak wash-out 3000 part deg over time}, which roughly matches our estimation for the final asymmetry. 

\subsection{Lower bound for $\mathbf{M_1}$}

As explained in the previous sections, the presence of many right-handed neutrinos will only help to increase the final baryon asymmetry (and hereby help to lower the lower bound on $M_1$) in the weak wash-out regime. Among the more closely considered scenarios, the one with the most decaying singlets will yield the largest final asymmetry.

Therefore combining eq.(\ref{eq absolute upper bound neff}) with eqs.(\ref{eq davidson ibarra}),(\ref{eq dilution factor}),  and (\ref{eq weak over time final B-L}), we find
\begin{equation}
	\eta _B \lesssim 
	  5 \cdot 10^{-8} \sqrt{K} \left(\frac {m_3}{0.05 \textrm{eV}} \right) 
	   \left(\frac {M_1}{10^9 \textrm{GeV}} \right) \, .
	 \end{equation}
With $K\approx 10^{-2}$, we find an approximate lower bound around of $T_r \approx M_1 \gtrsim 10^8$GeV. An increase of $K$ by an order of magnitude or so, would slightly increase the result of this formula. However, the numerics show that we would not be allowed to  neglect the wash-out processes during the decay anymore and that the final asymmetry already tends to be smaller than in the $K=10^{-2}$ case.

Numerically we were able to push the lower bound a little further as we were able reproduce the observed BAU with $K=10^{-2}$, $n\eff=10^4 $ and
\begin{equation}
	T_r=M_1=6.5 \cdot 10^{7} \textrm{GeV.}
	\end{equation}
We see that this relaxes the lower bound in eq.(\ref{eq stand lept low bound M1}) for standard thermal leptogenesis by approximately one and a half orders of magnitude. Moreover, the relaxation of the Davidson-Ibarra bound (as discussed in sec.\ref{sec davidson ibarra bound}) as well as a complete parameter scan might motivate a further relaxation of this bound.

\subsection{Upper mass bound for $\mathbf{\overline{m}^2}$} 
	\label{sec upper bound mbar many nu}
	The analysis in section \ref{sec upper bound mbar} that lead to the upper bound for $\overline{m}^2$ in the standard case is not valid in the presence of many singlets. The reason for this is that we cannot draw the conclusion that we are in the strong wash-out regime from the fact that we have degenerate light neutrinos as we did in the standard case. All we need now is that three singlets are in the strong wash-out regime. However, if we have many more singlets that are weakly coupled, they can determine the size of the final baryon asymmetry and not the strongly coupled ones that might determine the size of the light neutrino masses.
	
Combining eqs. (\ref{eq upper bound M1}), (\ref{eq upper bound m DI}), (\ref{eq dilution factor}), and (\ref{eq weak over time final B-L}) we find
\begin{equation} \label{eq upper mass barm many}
	 \left(  \frac {\overline{m}}{\textrm{eV}} \right)  
	 \lesssim  6
		 	 \, n^{\frac 1{12}}\eff
			\, K^{\frac 14} \, .
	\end{equation}
Thus, even if we imply the Davidson-Ibarra bound, we find that within our rough approximations the bound on the absolute neutrino mass scale becomes less stringent than bounds from other areas (e.g. \cite{Kraus:2004zw,Lobashev:1999tp,Hannestad:2003xv}). In scenarios that evade the Davidson-Ibarra bound, the upper mass bound on the light neutrino masses can be relaxed even further. Additionally, the bound in eq.(\ref{eq upper mass barm many}) seems rather conservative from the point that the $\Delta L=2$ scatterings do not need to be frozen out around $z=1$ in the weak wash-out regime but at lower temperatures.

\subsection{Further Issues} \label{sec further issues}

{\bf $\mathbf{\Delta L=2}$ Scatterings:}\\
So far we have neglected the possibility of $\Delta L=2$ scatterings. Our estimate for the corresponding bound, is the same as the one we used in the standard thermal leptogenesis case. As the Davidson-Ibarra bound, it does not get modified by the presence of many singlets, since the increase of the number of relevant diagrams is compensated by the increase of diagrams used for the see-saw mechanism. Again, we estimate the same reaction rate for the corresponding processes as in sec.\ref{sec complications} and therefore also use eq.(\ref{eq delta Leq2 up bound M1}) as a criterion for the validity of our considerations.

{\bf Quantum Effects:}\\
As already mentioned in our review of standard thermal leptogenesis in the previous section, there are also questions about the possible contributions of quantum effects. However, a detailed analysis of these effects in the presence of many singlets goes beyond the scope of this paper.
		
{\bf Initial Conditions:}\\		
While the initial conditions for the singlets abundances are not typically important in the strong wash-out regime, they are a crucial point in the weak wash-out regime, in which we are more interested. So far, we have always made the conservative assumption of zero initial abundance of the singlets. Without being quantitative let us make some comments on further possibilities.

If we assume to have a thermal initial abundance of the singlets (democratic reheating), we should also be able to drop the corresponding lower bound for the reheating temperature from refs. \cite{Buchmuller:2002rq,Giudice:2003jh} significantly, as a large number of singlet states would automatically imply a universe dominated by singlets whose decay produces a lepton asymmetry, cf. \cite{Giudice:2003jh}. However, compared to a universe dominated to by one singlet species we would probably not be able to lower the bound for the reheating temperature, since the only difference (in case of identical states) would be an increase of the wash-out rate, which might even slightly decrease the final asymmetry. Thus, while the setup also seems interesting for democratic reheating scenarios it seems less interesting for non-thermal baryogenesis scenarios, e.g. \cite{Lazarides:1991wu}.


\section{An Extra-Dimensional Example}\label{sec An Extra-Dimensional Example}

In this section we present a possible realization of the previously discussed leptogenesis scenario within a concrete model. Therefore, we adopt the frequently used idea that gauge singlets can propagate in an additional dimension, while all the standard model particles are confined to a 3-brane \cite{Arkani-Hamed:1998vp}. 
If there are no further interactions, one can formally integrate over this extra dimension and is left with a regular four dimensional theory with  a Kaluza-Klein tower of right-handed neutrinos. For a broad parameter space, this can lead to a quasi degeneracy of many of the Kaluza-Klein modes, which is a key ingredient needed for leptogenesis with many singlets, as we illustrated in the previous section.  

In this context, let us mention ref.\cite{Pilaftsis:1999jk}, which also  introduced an extra-dimensional model for a leptogenesis scenario that made use of the Kaluza-Klein tower of extra-dimensional neutrinos. However, the model is quite different from our proposal in many aspects. This is partially due to the different nature of the 5-d Majorana mass term, which is scalar-like in our model and vector-like in ref. \cite{Pilaftsis:1999jk}. Another difference is that ref.\cite{Pilaftsis:1999jk} needs to violate the orbifold symmetry to get a non-zero CP violation (since it only uses one generation of singlets), which is not necessary in our case. Also, we do not make use of any resonance effects that increase CP violation. Finally, since ref. \cite{Pilaftsis:1999jk} assumes much larger extra-dimensions, its leptogenesis scenario takes place after the electro-weak phase transition, which makes the quantitative treatment more difficult. The model from ref. \cite{Pilaftsis:1999jk} has been considered further in ref. \cite{Abada:2006yd}, which also suggests the consideration of three generations of singlets in the bulk to increase the final asymmetry.

Refs.\cite{Medina:2006hi,Gherghetta:2007au}, e.g., also consider extra-dimensional leptogenesis scenarios. While ref.\cite{Gherghetta:2007au} state that many decaying Kaluza-Klein states can increase the final asymmetry of their scenario, ref.\cite{Medina:2006hi} finds that a possibly decaying Kaluza-Klein tower of particles would not increase the final asymmetry in the respective model.

\subsection{The model}

Let us be more quantitative and use the treatment from refs. \cite{Arkani-Hamed:1998vp,Dienes:1998sb,Lukas:2000wn,Lukas:2000rg}, where the groundwork of this subsection has been developed.

As already mentioned, we use a five dimensional model, where the standard model particles are confined to the regular 3-brane. Using the notation of ref.\cite{Eisele:2006va} for this section, the corresponding part of the action is given by
\begin{equation} 
			S_\SM=\int d^4x\,dy\, {\cal L}_{SM}\,\delta (y) \, .
			\end{equation} 
Here, $y$ is the additional spatial degree of freedom, which we assume to be compactified on an orbifold $S_1/\mathbbm{Z}_2$ of radius $R$.

Additionally, we introduce three singlet fields $\Psi _i$ that can propagate in the extra-dimension and which transform as $P_5 \Psi _i=\gamma _5 \Psi _i$ under the parity transformation $P_5:y\rightarrow -y$. This allows us to write down the gauge and orbifold-symmetry invariant action
\begin{equation} 
			S_\bulk=\int d^4x\,dy\, \left[ 
			\overline{\Psi}_i i\gamma ^\alpha 
			\partial _\alpha \Psi _i
			-\frac 12 (M_{ij} \overline{\Psi ^c_i} \Psi _j
			+\hc )\right] \, ,
			\end{equation}
where $\Psi ^c_i$ is the corresponding charge conjugate field of $\Psi _i$ defined by
\begin{equation}
	\Psi ^c _i \equiv \left( \begin{array}{cc}
			0&\epsilon \\ \epsilon&0 \end{array} \right)
			\Psi_i^* \, ,
	\end{equation}
with $\epsilon=i\sigma _2$. We can always choose the $\Psi _i$ such that $M_{ij}=\diag (M_1,M_2,M_3)_{ij}$ with $M_i \, \epsilon \, \mathbbm{R}$ and $0\le M_1\le M_2 \le M_3$, which we will therefore assume in the following.

Further, we can introduce brane-bulk couplings that also respect the SM gauge invariance and the orbifold symmetry
\begin{equation} 
			S_\branebulk=\int d^4x\,dy\, \left[
			- \frac {g_{ij}}{\sqrt{M_5}}
			\overline{\Psi}_i P_L H 
			\left( (0,0),\ell_j^T \right)^T  
			+ \hc \right]
			\delta (y) \, ,
			\end{equation}
where the $g_{ij}$ are dimensionless Yukawa couplings and $M_5$ is a mass scale introduced for dimensional reasons.

The transformation properties of the $\Psi _i$ allow us to decompose them into modes
\begin{equation} \label{eq mode decompostion}
	\Psi _i (x,y)= \frac 1{\sqrt{\pi R}}
	\left( \begin{array}{r}
	\frac 1{\sqrt{2}} \Psi ^{(0)}_{R,i}(x) + \sum \limits^\infty _{n=1}
		\cos (ny/R) \, \Psi ^{(n)}_{R,i}(x) \\
		\sum \limits^\infty _{n=1}
		\sin (ny/R)\,  \Psi ^{(n)}_{L,i}(x) \end{array}
		\right),
	\end{equation}
with the respective right- and left-handed Weyl spinors $\Psi ^{(n)}_{R/L,i}(x)$.

We can now perform the integration over $y$ in the action and are left with a four dimensional theory with infinitely many particles due to the Kaluza-Klein tower of the $\Psi _i$. After the Higgs field develops a VEV $\langle \phi \rangle$ we find an infinite Majorana mass matrix for all the left-handed neutrino fields $(\nu,\hat \Psi ^{c(0)}_{R,i},\dots,\Psi ^{(n)}_{L,i},
\hat \Psi ^{c(n)}_{R,i},\dots)$ 
\begin{equation}\label{eq mass matrix gen}
	\left( \begin{array}{cccccc}
	0& - m^T_D/\sqrt{2} &\cdots& 0& - m^T_D&\cdots \\
	- m_D/\sqrt{2}& M&\cdots& 0& 0&\cdots \\
	\vdots&\vdots&\ddots&\vdots&\vdots&\ddots \\
	0&0&\cdots& -M&- n/R&\cdots \\
	- m_D&0&\cdots&-n/R& M&\cdots \\
	\vdots&\vdots&\ddots&\vdots&\vdots&\ddots 
	\end{array} \right) \, ,
	\end{equation}
with $\hat \Psi^{c(k)}_{R,i}\equiv -\epsilon [\Psi ^{(k)}_{R,i}]^*$ and  $m_{D,ij} \equiv g_{ij}\langle \phi \rangle/\sqrt{\pi R M_5}$.

If we further diagonalize the sub-matrices
\begin{equation}
	\left( \begin{array}{cc}
		 -M&- n/R\\
		 -n/R& M\end{array} \right)
	\end{equation}
by a change of basis $(\nu,\hat \Psi ^{c(0)}_{R,i},\dots,\Psi ^{(n)}_{L,i},
\hat \Psi ^{c(n)}_{R,i},\dots)\rightarrow (\nu,\hat \Psi ^{c(0)}_{R,i},\dots,\Psi ^{(n)'}_{-,i},
\hat \Psi ^{c(n)}_{+,i},\dots)$, the Majorana mass matrix becomes
\begin{equation}
	\left( \begin{array}{cccccc}
	0& - m^T_D/\sqrt{2} &\cdots& -m^T_{n,-}& -m^T_{n,+}&\cdots \\
	- m_D/\sqrt{2}& M&\cdots& 0& 0&\cdots \\
	\vdots&\vdots&\ddots&\vdots&\vdots&\ddots \\
	-m_{n,-}&0&\cdots& -\sqrt{M^2+n^2/R^2}&0&\cdots \\
	-m_{n,+}&0&\cdots&0& \sqrt{M^2+n^2/R^2}&\cdots \\
	\vdots&\vdots&\ddots&\vdots&\vdots&\ddots 
	\end{array} \right) \, ,
	\end{equation} 
where
\begin{equation}
	\left(m_{n,\pm}\right) =\pm \frac {m_{D,ij}}{\sqrt{2}} 
		\sqrt{1 \pm \frac{M_i}{\sqrt{M^2_i+n^2/R^2}}} \, 
	\end{equation}
and 
\begin{equation}
	\left(\sqrt{M^2+n^2/R^2} \right) _{ij}=\delta _{ij} 
		\sqrt{M_i^2+n^2/R^2}  \, .
	\end{equation}
A final change of basis with $\Psi ^{(n)'}_{-,i}\rightarrow \Psi ^{(n)}_{-,i} \equiv \I \, \Psi ^{(n)'}_{-,i}$ leads to%
\footnote{The author would like to thank Naoyuki Haba for related, private notes on the five dimensional see-saw (with vector-like masses) \cite{Haba:private} that inspired this treatment.}
\begin{equation}
	\left( \begin{array}{cccccc}
	0& - m^T_D/\sqrt{2} &\cdots& -\I \, m^T_{n,-}& -m^T_{n,+}&\cdots \\
	- m_D/\sqrt{2}& M&\cdots& 0& 0&\cdots \\
	\vdots&\vdots&\ddots&\vdots&\vdots&\ddots \\
	-\I \, m_{n,-}&0&\cdots& \sqrt{M^2+n^2/R^2}&0&\cdots \\
	-m_{n,+}&0&\cdots&0& \sqrt{M^2+n^2/R^2}&\cdots \\
	\vdots&\vdots&\ddots&\vdots&\vdots&\ddots 
	\end{array} \right) \, .
	\end{equation} 
We see that the Kaluza-Klein tower consists of pairs of fields with degenerate masses. Additionally, in case of $M_iR \gg 1$  a large number of pairs pairs will be quasi-degenerate with respect to their masses. Also, we note that the relative strengths and phases with which a mode couples to the SM neutrinos is the same for all the modes. Still, the couplings of the neutrinos to the different Weyl fields can differ by a constant factor and a possible overall phase.
	
Let us take a closer look at the Yukawa couplings of the various modes $\Psi ^{(n)}_{\pm,i}$, namely
\begin{eqnarray}
	\left( Y_{+,n} \right) _{ij }
	&\equiv &\left( \frac {m_{n,+}}{\langle \phi \rangle}\right) _{ij} 
	=+ \frac {g_{ij}}{\sqrt{2\pi R M_5}} 
		\sqrt{1 + \frac{M_i}{\sqrt{M^2_i+n^2/R^2}}} \, , \\
	\left( Y_{-,n} \right) _{ij }
	&\equiv &\left( \frac {\I \, m_{n,-}}{\langle \phi \rangle}\right) _{ij} 
	=- \I \frac {g_{ij}}{\sqrt{2\pi R M_5}} 
		\sqrt{1 - \frac{M_i}{\sqrt{M^2_i+n^2/R^2}}} \, ,
	\end{eqnarray}
for $n \le 1$. We see that for $n \gg M_iR$ the coupling strengths of the $\Psi ^{(n)}_{+,i}$ and $\Psi ^{(n)}_{-,i}$ modes become similar, whereas for $n \ll M_iR$ we find
\begin{eqnarray}
	\left( Y_{+,n} \right) _{ij } &\approx& 
		+ \frac {g_{ij}}{\sqrt{\pi R M_5}} 
		\left( 1-\frac 18  \left[ \frac {n}{M_iR} \right] ^2 \right) 
			\, , \label{eq Y+ approx} \\
	\left( Y_{-,n} \right) _{ij }&\approx& 
		- \frac {g_{ij}}{2 \sqrt{\pi R M_5}} \, \frac {n}{M_iR} 
		\label{eq Y- approx}\, ,
	\end{eqnarray}
from which we see that the couplings of the lower modes of $\Psi ^{(n)}_{-,i}$ can be strongly suppressed with respect to the lower modes $\Psi ^{(n)}_{+,i}$ if the corresponding Majorana mass scale $M_i$ is much larger than the compactification scale $R^{-1}$.

\subsection{CP asymmetry}

In the context of leptogenesis, it is of course important to determine the CP asymmetry produced by decaying modes of $\Psi _i$. With our considerations from previous chapters, this is now straightforward.%
\footnote{Since there are cancellations between the various contributions to the see-saw in this model, the perturbative approach using Feynman diagrams might lose its validity. However, if the cut-off for our extra-dimensional theory is around the scale $M_5$, this problems seems to be absent. \label{foot cut off}}

Applying eq.(\ref{eq CP asym many Ns}) we find
\begin{equation} 
	\varepsilon ^{(n)}_{\pm,i} = - \frac 3{16 \pi} \frac {M_{i,n}}
	{\langle \phi \rangle ^2}
		\frac 1{\left[(Y_{\pm,n}) (Y_{\pm,n})^\dagger \right]_{ii}} 
		{\textrm{Im}\left[ (Y_{\pm,n})  m_{\nu}^\dagger 
			(Y_{\pm,n})^T ) \right]_{ii}}  \, ,
	\end{equation}
with $M_{i,n}=\sqrt{M_i^2 +n^2/R^2}$. After further transformations we get
\begin{eqnarray} 
	\varepsilon ^{(n)}_{+,i}	&=& 
		- \frac 3{16 \pi} \frac {M_{i,n}}{\langle \phi \rangle ^2}
		\frac 1{\left[gg^\dagger \right]_{ii}} 
		{\textrm{Im}\left[ (g  m_{\nu}^\dagger 
			g^T ) \right]_{ii}} \\
	\varepsilon ^{(n)}_{-,i}	&=& 
		 \frac 3{16 \pi} \frac {M_{i,n}}{\langle \phi \rangle ^2}
		\frac 1{\left[gg^\dagger \right]_{ii}} 
		{\textrm{Im}\left[ (g  m_{\nu}^\dagger 
			g^T ) \right]_{ii}} = - \varepsilon ^{(n)}_{+,i} \, .
		\end{eqnarray}
Thus, each mode features two particle species, which yield exactly opposite CP asymmetries. Further the CP asymmetries only depend on the five dimensional wave number through the proportionality factor $M_{i,n}$. 

Let us also stress that in the case $M_3 \ge M_2 \gg M_1 \gg R^{-1}$, in which we have many quasi-degenerate neutrinos, we do not get any resonance or higher-order effects as discussed earlier, since all the quasi-degenerate particles with masses around $M_1$ have the same Yukawa couplings (except for a purely real or imaginary overall factor). 

Finally, we note that all singlets of one generation have the same Yukawa couplings (up to an overall factor). This leads to the fact that the effective light neutrino mass matrix can be written as the product of $3\times 3$ matrices instead of $n_N \times 3$. E.g. for $M_iR \gg 1$, the neutrino mass matrix in our model is given by  \cite{Arkani-Hamed:1998vp,Dienes:1998sb,Lukas:2000wn,Lukas:2000rg} (see also \cite{Haba:2006gt,Eisele:2006va})
\begin{equation} \label{eq simple 5d seesaw}
	m_\nu= m_D^T  m_D\frac {\pi R}2 \, .
	\end{equation} 
In this case the Casas-Ibarra parametrization \cite{Casas:2001sr} of the Yukawa couplings can be applied (with a possible additional overall factor and the heavy masses being replaced by $2/(\pi R)$). After some easy algebra with eq.(\ref{eq CP asym many Ns}) one then finds that the more stringent Davidson-Ibarra bound given in eq.(\ref{eq davidson ibarra}) is still valid in this scenario.

\subsection{Leptogenesis}

In addition to the masses of the various modes and the CP asymmetries they yield, the decay rates of the modes of the first generation are also needed to determine the final baryon asymmetry. Similarly to the previous chapters it is useful to characterize this quantity in terms of the parameter 
\begin{equation}
	K_{n,\pm} \equiv \frac{ \Gamma _{D,n,\pm}(T \rightarrow 0)}{H(T=M_1)}
	= \frac {\left[(Y_{\pm,n}) (Y_{\pm,n})^\dagger \right]_{11}
	M_{1,n} /(8\pi)}{1.66 g_* M_1^2/\MPl }
	= \left( 1 \pm \frac {M_1}{M_{1,n}} \right) K_0
	\, ,
	\end{equation}
for $n\ge 1$ with a corresponding definition for $K_0$ 

Considering eq.(\ref{eq simple 5d seesaw}) and the four-dimensional effective neutrino mass (cf. eq(\ref{eq eff nu mass 4d})), one can define the five-dimensional effective neutrino mass as \begin{equation}
	\tilde m_{\nu, \textrm{5d}}= (m_D m_D^\dagger)_{11} \frac {\pi R}2. 
	\end{equation} 
Yet, this yields the relation
\begin{equation}
	K_0 = \frac {\tilde m_{\nu, \textrm{5d}}}{m_*} \frac 2{\pi M_1R} \, ,
	\end{equation}
which implies that the parameter $K_0$ is pushed to smaller values with respect to the corresponding four dimensional scenario. Hence, while the size of the measured neutrino mass differences might make $K \gg 1$ more plausible in four dimensions, the weak wash-out regime can be more natural in this scenario. 

Once we have fixed the values $M_1$, $R$, $T_R$, and $K_0$ we can already use eq.(\ref{eq bar as approx weak many N}) and (\ref{eq weak over time final B-L}) to approximate the final $B-L$ asymmetry of this leptogenesis setup. The key feature of this model is the fact that modes that yield opposite CP asymmetries couple with different strengths to the lepton doublets. Being in the weak wash-out regime, their contributions to the final $B-L$ asymmetry differ in size, which prevents them from canceling each other. If the reheating temperature equals the mass of the lightest of the heavy neutrinos ($T_R=M_1$), there are of the order of $M_1R$ quasi-degenerate neutrinos which will reach significant particle abundances. Thus, if we assume maximal CP violation, a rough estimate for the final $B-L$ asymmetry is 
\begin{equation} \label{eq final asym extra rad}
	N_{B-L}  \approx 
	  10^{-7}\left(\frac {m_3}{0.05 \textrm{eV}} \right) 
	   \left(\frac {M_1}{10^9 \textrm{GeV}} \right)
	   (M_1R) \,  K_0 \, ,
	 \end{equation}
or
\begin{equation} \label{eq final asym extra mat}
	N_{B-L}  \approx 
	  2.5 \cdot 10^{-6}\left(\frac {m_3}{0.05 \textrm{eV}} \right) 
	   \left(\frac {M_1}{10^9 \textrm{GeV}} \right)
	   (M_1R)^{\frac 14}  \, K_0^{\frac 34} \, 
	 \end{equation}
depending on the fact if the singlets come to dominate the universe or not, and therefore on the size of $M_1R$ (cf. eq.(\ref{eq upper bound n})).

Using eq.(\ref{eq dilution factor}) we can see that values such as $M_1=2\cdot 10^8$ GeV, $R^{-1}=10^5$ GeV, and $K_0=10^{-2}$ already seem to yield a baryon asymmetry of the right order of magnitude. 

Let us be more precise. If equation (\ref{eq bar as approx weak many N}) and therefore also eq.(\ref{eq final asym extra rad}) is valid, we see that the contribution of a pair of particles $\Psi _{\pm,1}^{(n)}$ to the final baryon asymmetry is independent of their actual masses, since $K_{+,1}^{(n)}-K_{-,1}^{(n)}=2 M_1/M_{1,n}$, which means the $M_{1,n}$ in the denominator will cancel with the $M_{1,n}$ in $\varepsilon _1$. Thus, all pairs with masses low enough that there will be a significant abundance of them approximately contribute the same amount to the final asymmetry.

\begin{figure}
	\begin{center}\includegraphics[%
	  width=0.45\columnwidth,keepaspectratio]
	  {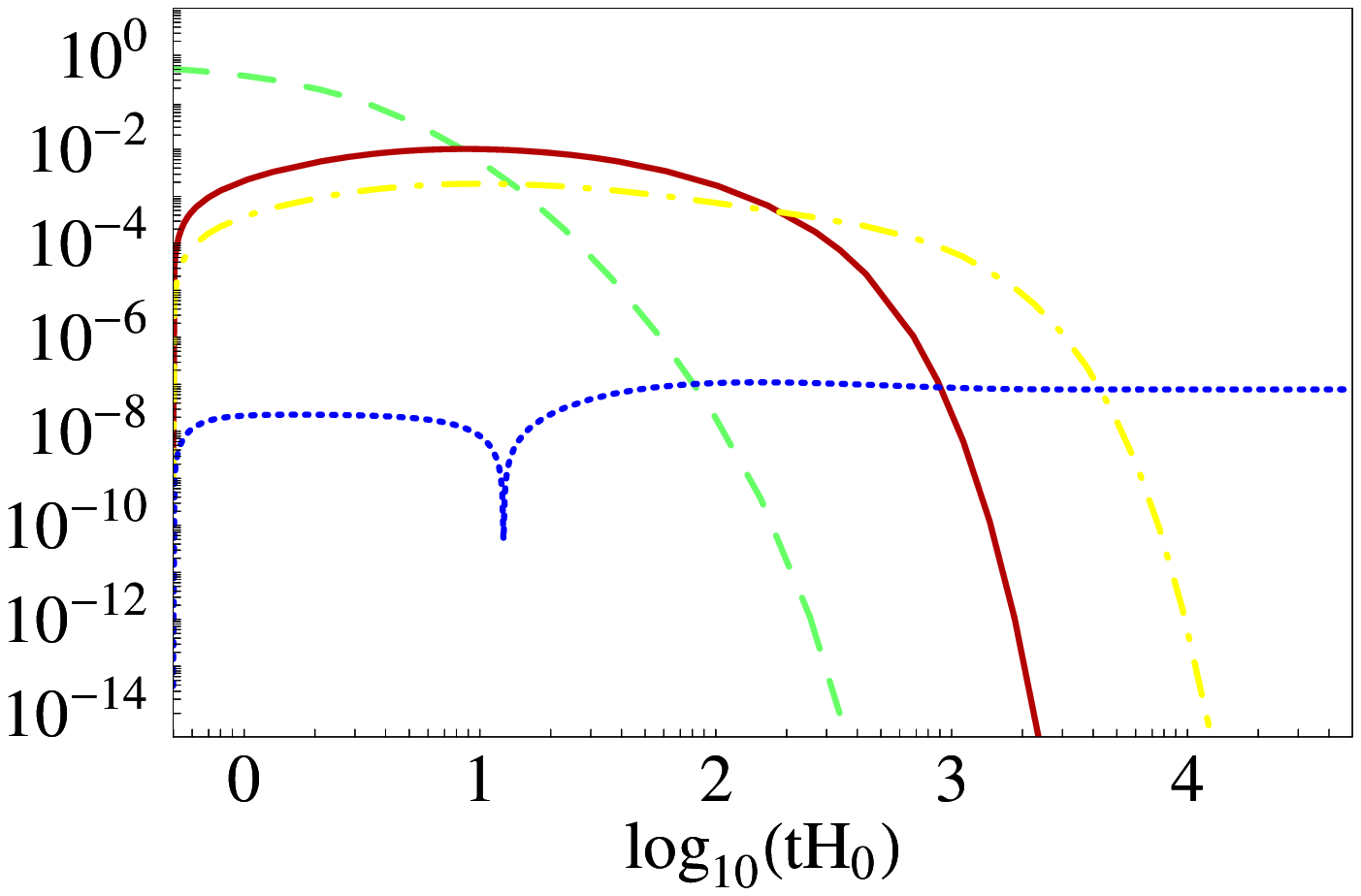}
	  \includegraphics[%
	  width=0.45\columnwidth,keepaspectratio]
	  {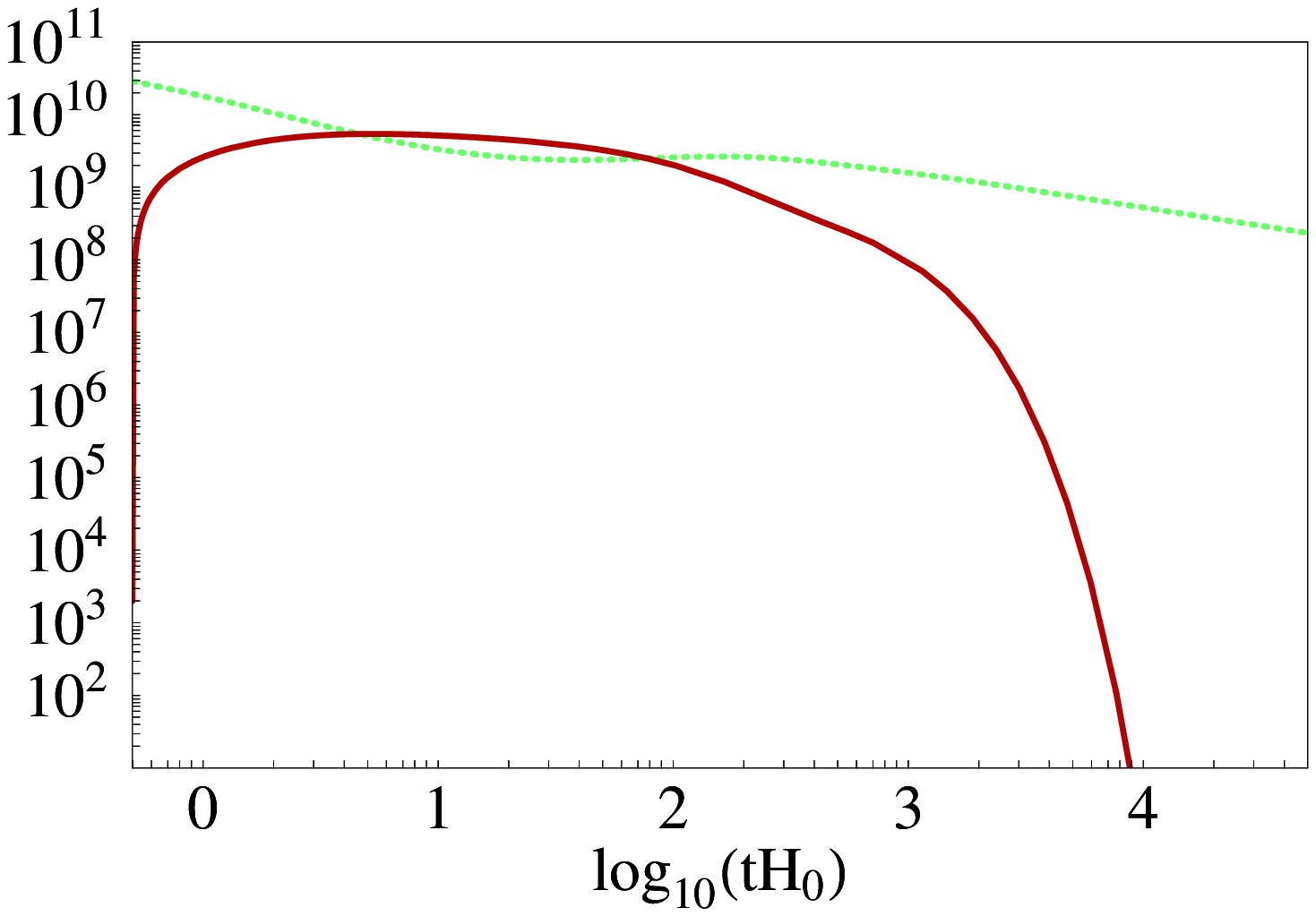}
	  
	  \end{center}
	\caption{\label{fig weak wash-out 2000 xtra dim}\textit{The two graphs show the numerical solution for $M_1R$ pairs of particles with parameter values as described in the text. In the graph on the left, the green/dashed line represents the equilibrium abundance of each of the particle species per photon. The red/solid line shows the corresponding value for a ``+''-mode, while the behavior of the corresponding ``-''-mode is illustrated by the yellow/dashed-dotted line. The absolute amount of the $B-L$ asymmetry is again represented by the blue/dotted line and reaches a final value of $7.8 \cdot 10^{-8}$. In the second graph the behavior of the respective energies in a comoving volume is shown, where the red/solid line corresponds to the singlets and the green/dotted line to radiation. The comoving volume contains one photon at early times.}}
	\end{figure}

If equation (\ref{eq weak over time final B-L}) and therefore also eq.(\ref{eq final asym extra mat}) is valid, we need to consider $(K_{+,1}^{(n)})^{3/4}-(K_{-,1}^{(n)})^{3/4}$, which means that pairs with larger masses will contribute slightly less, even if they get produced in a significant abundance. This means, instead of treating all $M_1R$ particle pairs as different, we can work with a system of completely similar particle pairs of mass $M_{1,n}$ to get an approximate lower bound for $M_1$ and the reheating temperature. This has been done in \mbox{figure \ref{fig weak wash-out 2000 xtra dim}} numerically, where we were able to produce a $B-L$ asymmetry that was even slightly larger than the observed value with $R^{-1}=10^5$ GeV, $K_0=10^{-2}$, and
\begin{equation}
	T_r =M_1=2 \cdot 10^8 \textrm{GeV,}
	\end{equation}
while we also assumed maximal CP asymmetry.

Of course, lower values would be possible if we would consider also $R^{-1}< 10^5$GeV. However, due to our arguments in the next subsection this does not seem to be motivated.

\subsection{Constraints From Extra-Dimensional Gravitons}

In addition to the open points listed in section \ref{sec further issues}, this extra-dimensional scenario has to consider the effects of gravitons that are likely to propagate in the extra dimension. In particular, we need to ensure that excitations of gravitational modes do not spoil the leptogenesis scenario by producing too much entropy, especially during BBN.
Therefore, we consider the possible production and decay of gravitational modes, mainly using the analysis from ref.\cite{Arkani-Hamed:1998nn}.

Since the coupling constant of a 5-d graviton to the standard model particles is proportional to $1/M_5^{3/2}$, a single mode coupling is proportional to $1/(M_5^{3} R)^{1/2}$ (cf. eq.(\ref{eq mode decompostion})). By dimensional analysis the decay rate of a single Fourier mode of mass $M_\Gr$ in its rest frame (from the four dimensional point of view) is therefore
\begin{equation}\label{eq gr dec simple}
	\Gamma _{D, \Gr} \approx n_{\textrm{dc}}\frac {M_\Gr^3}{M_5^3 R}
	= n_{\textrm{dc}}\frac {M_\Gr^3}{\MPl^2},
	\end{equation}
where $n_{\textrm{dc}}$ is the number of decay channels.

If we define the parameter $K_\Gr$ for a graviton mode corresponding to eq.(\ref{eq K def}), we get 
\begin{equation}
	K_\Gr \equiv \frac {\Gamma _{D, \Gr}}{H(T=M_\Gr)}
		\approx \frac {n_{\textrm{dc}}}{g_*^{1/2}} \frac {M_\Gr}{\MPl} \, .
	\end{equation}	
The typical decay time of these modes is $\Gamma _\Gr^{-1} \equiv K_\Gr^{-1} H(T=M_\Gr)^{-1}$ (cf. eq.(\ref{eq time od decay})), which corresponds to the temperature 
\begin{eqnarray}
	T_\Gr &=& M_\Gr \left( \frac {a(t=0)}{a(t=\Gamma _\Gr^{-1})}  \right)
		= M_\Gr (2K^{-1}_\Gr)^{-\frac 12} \\
		&\approx & \left( 2\frac {n_{\textrm{dc}}}{g_*^{1/2}} \right)^{\frac 12} 
			\left( \frac {M^3_\Gr}{\MPl} \right)^{\frac 12}
				\label{eq TGr ex}  \\
		&\gtrsim &\left( \frac {M^3_\Gr}{\MPl} \right)^{\frac 12}
			\label{eq TGr} \\
		&\approx & \left( \frac {M_\Gr}
			{10 \textrm{ TeV}} \right)^{\frac 32}
			\cdot 0.3 \textrm{ MeV.}
 	\end{eqnarray}
where we estimated $n_{dc}\gtrsim g_*$ in eq.(\ref{eq TGr}).

Since the lowest mass of a gravitational mode (other than the zero mode) approximately equals the compactification scale, we find the requirement that the compactification scale approximately needs to be of the order 50 TeV or larger. Otherwise, the gravitational modes might decay late and change the predictions of big bang nucleosynthesis.

Even if the gravitons decay early enough to avoid bounds from BBN, it is also important that they do not get to dominate the energy density of the universe before that time. Otherwise the entropy production during their decay could invalidate our estimations for the size of the produced baryon asymmetry. 

To consider this possible issue, we can use eq.(\ref{eq no dens many nu}) to roughly estimate the abundance of a gravitational mode in a comoving volume, which contains one photon. Since the number of modes with mass $M_\Gr$ is approximately $M_\Gr R$ we can estimate the energy per comoving volume stored in gravitons of mass $M_\Gr$ at the time of their decay by
\begin{equation}
	E_\Gr \approx (M_\Gr R) \, K_\Gr \, M_\Gr \approx 
		\frac {n_{dc}}{g_*^{1/2}} \frac {M_\Gr^3 R}{\MPl} \, ,
	\end{equation}
whereas the radiation energy stored in the same volume at the same time is approximately  ${g_*} T_\Gr /2 $ (cf. eq.(\ref{eq TGr})). 

The ratio of the two energies is therefore approximately
\begin{equation}
	\frac {E_\Gr}{E_\textrm{rad}} \approx 0.02 \cdot M_\Gr R \sqrt{ \frac {M_\Gr}{\MPl}} 
	\, ,
	\end{equation}
which indicates that even for a compactification scale as low as 10 TeV, the reheating temperature can be as high as $10^{10}$ GeV without a significant entropy production due to gravitons.

Thus, within the limits of our estimations the lower limit on the reheating temperature for the leptogenesis scenario considered in this section does not get additional bounds from the consideration of gravitational modes other than a compactification scale in the 50 TeV range or larger. 


\section{Summary and Outlook}

Considering leptogenesis scenarios with many singlet neutrinos, we have found that it is possible to lower the reheating temperature of standard thermal leptogenesis by approximately one and a half orders of magnitude. This effect can be achieved in the weak wash-out regime, while the strong wash-out scenarios become slightly more restrictive compared to the standard case. 
We also argued that the weak wash-out scenario can become more natural in theories with many singlets. Approximative formulae have been given that help to illustrate the effect of the additional particles. Additionally, we have checked our approximations by comparing them to results from numerical solutions of the corresponding Boltzmann equations. 

We further argued that the bound on the absolute neutrino mass scale from standard thermal leptogenesis can get weakened in the considered scenarios and might fall behind bounds from other areas of particle and astro-particle physics.

As a specific realization of a leptogenesis with many singlets, we considered an extra-dimensional scenario, where the Kaluza-Klein excitations of right-handed singlets serve as right-handed neutrinos in the weak wash-out regime and provide all necessary ingredients for leptogenesis below the usual bounds.
We also considered the excitations of gravitational bounds for this model.
Within our approximations we find that these bounds still allow for a significant reduction of the necessary reheating temperature.

Neither for the general case of many neutrinos, nor for the explicit extra-dimensional model we have explicitly performed parameter scans. This means, there might be a parameter range for which the bound for the reheating temperature can even be lowered further.

The fact that the Davidson-Ibarra bound does not generally hold in the case of many singlets might also leave room for a further relaxation of the lower bound for the reheating temperature. This is especially interesting in the context of our arguments for the lift of the upper mass bound for the light neutrinos masses.

An interesting topic that goes beyond the scope of this paper is the inclusion of flavor dependence in this setup, especially, since it may be that the interference of tree-level and the self-energy correction diagrams for the CP asymmetry gives important additional effects, if the particles in the loop propagate in the opposite direction compared to fig.\ref{fig lept graphs}.

\section{Acknowledgments}

I would especially like to thank Yosef Nir for valuable comments on the first version of this paper, in particular in the context of the Davidson-Ibarra bound with many neutrinos.

Further, I would like to thank Michael Ratz for numerous fruitful discussions that had a large impact on the quality of the present work, as well as for valuable comments on the manuscript. I would also like to thank Naoyuki Haba for introducing me to the idea of extra-dimensional neutrinos, discussions on the 5d see-saw and leptogenesis at early stages, as well as for private notes on the 5d see-saw. Moreover, my thanks go to Michael Pl\"umacher and Oleg Lebedev for useful discussions.
Thanks for financial support go to the ``Graduiertenkolleg
1054'' and the ``SFB-transregio 27'' of the ``Deutsche Forschungsgemeinschaft'' as well as the cluster of excellence ``Origin and Structure of the Universe''.

\newpage
\begin{appendix}
\section{Reaction Rates} \label{sec app reac rates}
For our numerical solutions we used the reaction rates as mainly listed in ref.\cite{Buchmuller:2004nz}.

The decay rate is given by \cite{Kolb:1979qa}
\begin{equation}
	D \equiv \frac {\Gamma _D}{Hz} = K \, z \, \frac {K_1(z)}{K_2(z)} \, ,
	\end{equation}
where the $K_i(z)$ are modified Bessel functions of the second kind.

The $\Delta L=1$ scattering rates for the s- and t-channel processes are given by \cite{Plumacher:1997ru}
\begin{eqnarray}
	S_s &\equiv & \frac {\Gamma _{\Delta L=1,s}}{Hz} 
	= 2 \cdot \frac {K_S}{12} f_s(z) \, \\
	S_t &\equiv & \frac {\Gamma _{\Delta L=1,t}}{Hz} 
	= 2 \cdot 2 \cdot \frac {K_S}{12} f_t (z) \,
	\end{eqnarray}
with 
\begin{equation}
	K_S \equiv \frac {\tilde m_1}{m_*^S}\, , \quad
	{m_*^S} \equiv \frac {8 \pi ^2}9  \frac{m_t^2}{v^2} m_* \approx 10 m_*
	\end{equation}
and
\begin{equation}
	 f_{s/t}(z)  \equiv \frac {\int_{z^2}^\infty d\psi g_{s/t}(\psi /z^2)
		\sqrt{\psi} K_1(\sqrt{\psi})}{z^2K_2(z)} 
	\end{equation}
and
\begin{eqnarray}
	g_s(x) &\equiv & \left( \frac {x-1}{x} \right)^2 \\
	g_t(x) &\equiv & \frac {x-1}x \left[ \frac{x-2+2a_h}{x-1+a_h}
		+\frac {1-2a_h}{x-1} \log \left( \frac {x-1+a_h}{a_h}
		\right) \right] \, ,
	\end{eqnarray}
where we had to introduce a Higgs mass ($a_h\equiv(m_h/M_1)^2$) to regularize the infrared divergence of the t-channel diagrams but treated the Higgs boson and all other particles (except for the right-handed neutrinos) as massless everywhere else .

To cut down the time needed for the numerical calculations we only calculated $f_{s/t}(z)$ at a fixed number of points (20-30) and then linearly extrapolated the functions.

Further, the wash-out rate is given by
\begin{equation}
	W= \frac {\Gamma _W}{Hz}=W_{ID}+W_{\Delta L=1}
	\end{equation}
with
\begin{equation}
	W_{ID}\equiv \frac 12 D \frac {N_{N_1}\eq}{N_\ell\eq}
		=\frac 14 K z^3 K_1(z)
	\end{equation}
and
\begin{equation}
	W_{\Delta L=1} = S_t \frac {N_{N_1}\eq}{N_\ell\eq}
		+ S_s \frac {N_{N_1}}{N_\ell\eq} \, ,
	\end{equation}
where $N_{N_1}$ is the actual abundance of the right-handed neutrinos in a comoving volume containing one photon, ${N_{N_1}\eq}=3 z^2 K_2(z)/8$ is the corresponding value for an equilibrium distribution, and $N_\ell\eq=3/4$ is the corresponding value for massless leptons.

The generalization to the case of many decaying the neutrinos is simple (cf. eqs. (\ref{eq boltzmann many N}) and (\ref{eq boltzmann many B-L})), since all one has to do is to do the interchange $M_1 \rightarrow M_n$ and $K \rightarrow K_n$ for each singlet state.\footnote{%
It is important to note that, aside from the change due to the Yukawa couplings, $K_n$ acquires a factor of $M_n/M_1$ and not of the inverse, since only the $M_1$ factor in the decay rate gets modified while $H_0$ does, of course, not change.}
 However, one has to remember that the factor in the denominator $Hz$  does obviously not change for each state. To find the correct rates one has to make the transformations
\begin{equation}
	D \rightarrow D_n: D(z) \rightarrow D\left(  
		z\frac {M_n}{M_1} \right) \frac {M_1}{M_n}
	\end{equation}
and in the same way for the $S$ and $W$ terms (also $a_h\rightarrow a_h (M_1/M_n)^2$).

With these considerations it is also simple to get the rates for the system described by eqs.(\ref{eq boltzmann many N over t}) to (\ref{eq T in scen over time}). Namely,
\begin{equation}
	\frac {\Gamma _{D,n}(M_n,T)}{H_0}=D_n (M_n,T)\, \frac T{M_1}
	\end{equation}
and accordingly for $S$ and $W$.

\section{The Davidson-Ibarra bound revisited} \label{sec app DI}

In this part of the Appendix we generalize the treatment from ref.\cite{Davidson:2002qv} to find an upper bound for the CP-asymmetry for a setup with $n_N$ heavy singlets and three light neutrinos.
In this case, one can generalize the Casas-Ibarra parametrization \cite{Casas:2001sr} of the Yukawa couplings to
\begin{equation}
	g= \frac 1{\langle \phi \rangle ^2} D_{\sqrt{M}} R 
	D_{\sqrt{m}} U^\dagger \, ,
	\end{equation}
where $D_{\sqrt{M}}$ and $D_{\sqrt{m}}$ are diagonal matrices that contain the masses of the heavy and light neutrinos, respectively. Here, we work in the basis where the heavy mass matrix is diagonal and $U$ is the mixing matrix of the light neutrinos. $R$ is a complex $n_N \times 3$ matrix that fulfills the condition 
\begin{equation}
	R^T R = \mathbbm{1}_{3\times 3}.
	\end{equation}
Let us use a mathematical trick and introduce $n_N-3$ additional light neutrinos with zero couplings. We can easily do this, since these particles do not show up in any particle physics experiment. Now, the Yukawa couplings can be parametrized by 
\begin{equation}
	g= \frac 1{\langle \phi \rangle ^2} D_{\sqrt{M}} R' 
	D'_{\sqrt{m}} U'^\dagger \, ,
	\end{equation}
with $D'_{\sqrt{m}}= \textrm{diag}(m_1,m_2,m_3,0,\dots,0)$ and
\begin{equation}
	U'\equiv \left( \begin{array}{cc}
			U & 0 \\
			0 & \mathbbm{1}_{(n_N-3)\times (n_N-3)} \end{array}
			\right).
	\end{equation}
$R'$ is now an arbitrary complex orthogonal $n_N\times n_N$ matrix with
\begin{equation}
	R'^T R' = \mathbbm{1}_{n_N\times n_N}.
	\end{equation}
As in the $3\times 3$ case \cite{Davidson:2002qv} eq.(\ref{eq CP asym many Ns}) can now be transformed to
\begin{equation}
	\varepsilon _i \approx 
		\frac 3{16 \pi} \frac {M_i}{\langle \phi \rangle ^2}
		\frac {\sum _j m_j^2 \, \textrm{Im}(R'^2_{1_j})} 
		{\sum _j m_j |R'_{1_j}|^2} \, ,
	\end{equation}
with $m_k=0$ for $k>3$. 

Using $\sum _j R'^2_{1j}=1$ we then get
\begin{equation}
	|\varepsilon _i | \lesssim 
	\frac 3{16 \pi} \frac {M_i}{\langle \phi \rangle ^2}
		\frac{ \textrm{Max}(\{m_j^2\})
		- \textrm{Min}(\{m_j^2\})}
		{ \textrm{Max}(\{m_j\})
		+ \textrm{Min}(\{m_j\})}\\
		\, = \, 
	\frac 3{16 \pi} \frac {M_i}{\langle \phi \rangle ^2}
		\, m_3 \, ,
	\end{equation}
where $m_3$ is essentially a free parameter.

\end{appendix}


\clearpage
\newpage
\bibliographystyle{OurBibTeX}
\bibliography{references}

\end{document}